\documentclass[]{article}
\usepackage{times}
\usepackage{makeidx}  
\usepackage{microtype}
\usepackage{graphicx}
\usepackage{enumitem}
\usepackage{multirow}
\usepackage{longtable} \usepackage{afterpage} 
\usepackage{tabularx}
\usepackage{url}
\usepackage{multirow}

\usepackage{rotating}

\usepackage{geometry}
\newcommand{\poubelle}[1]{} 

\newcommand{\xc}[1]{\textsc{#1}}

\newcommand{\xb}[1]{\textbf{#1}}

\newcommand{\zi}[1]{\textit{#1}}

\newcommand{\titleSoS}[1]{\noindent\xb{#1}}
\newcommand{\titleITILprocess}[2]{\vspace{-1.8mm}\begin{itemize} \item \xb{#1}: #2 \end{itemize}\vspace{-1.5mm}}
\newcommand{\tabIO}[1]{\noindent \small #1 \normalsize \vspace{0.1mm}}
\def\tabColUn{8cm} \def\tabColDeux{5.1cm} 

\begin{document}


\title{Aligning a Service Provisioning Model of a Service-Oriented System with the ITIL v.$3$ Life Cycle\\\zi{Technical Paper}\footnote{A shorter version of this work has been submitted to the Fifth International Conference on Exploring Service Science (\xc{iess} $1.5$).}}
\author{Bertrand Verlaine \and Ivan J. Jureta \and St\'ephane Faulkner
\\PReCISE Research Center \& Business Administration Department\\University of Namur, Belgium
\\\{bertrand.verlaine,ivan.jureta,stephane.faulkner\}@unamur.be}
\date{\today}
\maketitle

\begin{abstract}
Bringing together the \xc{ict} and the business layer of a service-oriented system (\xc{s}o\xc{s}) remains a great challenge. Few papers tackle the management of \xc{s}o\xc{s} from the business and organizational point of view. One solution is to use the well-known \xc{itil v.}3 framework. The latter enables to transform the organization into a service-oriented organizational which focuses on the value provided to the service customers. In this paper, we align the steps of the service provisioning model with the \xc{itil v.}3 processes. The alignment proposed should help organizations and \xc{it} teams to integrate their \xc{ict} layer, represented by the \xc{s}o\xc{s}, and their business layer, represented by \xc{itil v.}3. One main advantage of this combined use of \xc{itil} and a \xc{s}o\xc{s} is the full service orientation of the company.

\providecommand{\keywords}[1]{\vspace{0.3cm}\noindent\textbf{Key words: } #1}
\keywords{Enterprise Architecture, IT Service Management, ITIL v3, Service-Oriented Paradigm, Service Provisioning}
\end{abstract}

\section{Introduction}
The current management approaches of the information and communication technologies (\xc{ict}) increasingly focus on a global view, often called the \zi{enterprise architecture}~\cite{ieee1471_2000,EAwinterFischer,ross2006enterprise}. The latter brings together the \xc{ict} layer and the business layer of organizations. The \xc{ict} level consists of the technical infrastructure and the application components, making up the Information System (\xc{is}) which supports the organizational activities. Concerning the business level, it consists of the vision and strategy of the organization, its organizational structure and its business processes.
\\With regard to the \xc{ict} level, and more specifically the software architecture of \xc{is}s, one of the last significant evolutions is the service-oriented paradigm. Although its definition varies somewhat in the literature, we refer to it as ``a paradigm for organizing and utilizing distributed capabilities that may be under the control of different ownership domains. It provides a uniform means to offer, discover, interact with and use capabilities to produce desired effects consistent with measurable preconditions and expectations''~\cite{OASISSOA}. This paradigm leads to the implementation and management of Service-oriented Systems (\xc{s}o\xc{s}). The latter offers the means, i.e., the software components and, to a lesser extend, the hardware components, to manage distributed and independent software functionalities named services ---services are the software building block used in a \xc{s}o\xc{s} to perform the functionalities required by the stakeholders~\cite{DBLP:journals/internet/HuhnsS05}.  
\\The transition from a traditional \xc{is} towards a \xc{s}o\xc{s} is a strategic and crucial decision in organizations. Making this transition a success requires to apply some of the existing Service-oriented Software Engineering (\xc{sose}) models and best practices ---\xc{sose} being the process followed to develop a \xc{s}o\xc{s}. As with the Traditional Software Engineering (\xc{tse}), \xc{sose} is the application of a systematic and structured approach for the analysis, the design, the conception, the implementation, the operation and the maintenance of \xc{s}o\xc{s}s. Unlike \xc{tse}, \xc{sose} has, on top of that, to organize the creation, the publication, the discovery, the composition, the evaluation and the monitoring of services~\cite{BeyondSOSE}. Differences between \xc{tse} and \xc{sose} --e.g., different system architectures, consumer/provider relationship, the importance of trust, the non-functional requirements variability and management, the heterogeneity acceptance, and so on~\cite{DBLP:conf/wicsa/GuL09}-- justify the need for specific and/or adapted models, methodologies and tools in order to create and manage \xc{s}o\xc{s}.
\\However, most of the recent research works about the \xc{sose} topic put aside the organizational and management aspects: ``the existing \xc{sose} methodologies focus mainly on the design and analysis part of the \xc{sose} process, but pay little or not sufficient attention to the constructing, delivering and management part''~\cite{DBLP:journals/soca/GuL11}. The integration of the service orientation and the service management, which becomes the foundations of the enterprise architecture~\cite{DBLP:conf/sac/BraunW07}, should help the software engineering teams and the organization leaders in their daily work. This should also improve and enhance the relationships between them because of a better understanding of the work and responsibilities of each one. This evolution is named the \zi{true Service-oriented Architecture} by Engels et al.~\cite{DBLP:conf/iceis/EngelsHHJLRVW08}. It is characterized by considering the business and managerial aspects when the services are created and provisioned.

\subsection{Motivations \label{motivations}}
Most of the approaches proposed to manage \xc{s}o\xc{s}s lack of relations with organizational models, whereas these relations are needed for a in-deep understanding of the service-oriented paradigm and of the specificities of the \xc{s}o\xc{s} management~\cite{TowardsEnhancedSOA}. The main objective is to align the management of the \xc{s}o\xc{s} with the management of the organization that uses the \xc{s}o\xc{s} to support its business processes. In the literature, some works tackle this issue by focusing on the governance issue (see the related work~\S\ref{relatedWork} for more details). We argue that a global governance is needed but that a process alignment can help the \xc{ict} teams in charge of the \xc{s}o\xc{s} to coordinate their work with the rest of the organization, and conversely.

We chose to use the Information Technology Infrastructure Library (\xc{itil}) as the reference organizational framework~\cite{itilSS,itilSD,itilST,itilSO,itilCSI}. The two main reasons are the significant attention paid to the notion of service and its usual use in many \xc{ict} companies. \xc{itil} is a collection of best practices for the management of organizations providing \xc{it} services. \xc{itil} provides the guidelines to organize a set of organizational capabilities --i.e., ``an appropriate mix of people, process and information technologies''~\cite{itilSS}-- for providing value to internal and/or external consumers in the form of \xc{it} services. Regarding \xc{itil}, its last version clearly advises the \xc{it} industry to use the service-oriented paradigm in order to structure \xc{is}s: ``It is strongly recommended that business processes and solutions should be designed and developed using a service-oriented architecture approach''~\cite[\S3.10]{itilSD}. However, \xc{itil} books and community do not explain how to develop this service-oriented architecture (\xc{soa}) neither detail the steps to follow in order to provision this system with new services. Moreover, the links between the life cycle, phases and processes of \xc{itil v.}3 and the steps of the implementation and the development of \xc{s}o\xc{s} are not defined neither explained. Some initiatives in the scientific literature tackle the links of the service-oriented paradigm and \xc{itil}, but these works only focus on the governance issue (e.g.,~\cite{DBLP:conf/iceei/SusantiS11,mapping6phaseSOAandITIL} which are further described in~\S\ref{relatedWork}).

\subsection{Contributions}
The main contribution of this paper lies in the alignment of the \xc{itil v.}3 processes with the steps of a \xc{s}o\xc{s} implementation methodology. To do so, we first detail the service-oriented development methodology of Papazoglou~\cite{DBLP:journals/ijwet/PapazoglouH06} that we named the \xc{s}o\xc{s} provisioning model. It is structured around the Spiral Model and covers the analysis, the design and the implementation of services in a \xc{s}o\xc{s}. Then, we define and explain the links between the steps of the \xc{s}o\xc{s} provisioning model and the processes of the \xc{itil v.}3 life cycle. The result is an alignment between the business layer represented by the \xc{itil} best practices and the \xc{ict} layer which corresponds to the \xc{s}o\xc{s} provisioning model.
\\These contributions should help companies that are organized --or wanting to get organized-- according to the \xc{itil} processes and working with a \xc{s}o\xc{s} to execute an enterprise architecture, i.e., choosing and then successfully developing a global organizational strategy.

From a more theoretical point of view, this work contributes to solve an open issue of \xc{itil v.}3: the use of an service-oriented architecture is highly recommended by \xc{itil} but the relations between them are not further explained\footnote{The five official books describing \xc{itil} contain more than $2 000$ pages. Only a paragraph (\S3.10 in \cite{itilSD}) is about these two topics and their (possible) relations.}.

\subsection{Organization}
This paper is organized in six sections besides this introduction. After the presentation of the \xc{itil v.}3 life cycle~(\S\ref{presentationITIL}), we compare a service-oriented development methodology with \xc{itil} in order to underline the gaps and issues raised by their combined use~(\S\ref{currentMethodology}). In order to make the alignment more accurate, we expand the service-oriented development methodology based on the Spiral Model~(\S\ref{SpiralModelRevisited}). In the next section~(\S\ref{AlignmentSoSprovisioningModel}), we develop and comment the links between the steps of the \xc{s}o\xc{s} provisioning model and \xc{itil v.}3. Then, we propose a validation framework to experiment our contributions in a real context~(\S\ref{validationFrameworkProposed}). Its main objective is to study the effects of the alignment proposed on an organization and on service implementation teams which use it in their daily work. This work ends with a discussion of the related work~(\S\ref{relatedWork}) followed by the conclusion and future work~(\S\ref{CCL}).

\section{A Short Presentation of ITIL v.$3$\label{presentationITIL}}
The \xc{itil} framework is an initiative of a British agency --the Office of Government Commerce-- in the late $80$'. A decade later, \xc{itil} has been clarified and improved within a second version. This collection of best practices for \xc{it} Service Management (\xc{itsm}) is probably the most known and used framework in high-tech companies. \xc{itsm} requires to focus the whole organization on the customers and, more specifically, on the value that they get by using its \xc{it} services~\cite{DBLP:conf/eee/HochsteinZB05}. An \xc{it} service is a combination of information technologies, people and processes used to provide directly or indirectly a service~\cite{itilSS}. The latter is ``a means of delivering value to customers by facilitating outcomes customers want to achieve without the ownership of specific costs and risks''~\cite{itilSS}.
\\The third version of \xc{itil} has been released in $2007$ and revised\footnote{From here, each time we mention \xc{itil v.}3, we refer to its last version, namely the one revised in $2011$.} in $2011$. A recent empirical study~\cite{DBLP:conf/ecis/MarroneK10} demonstrates that \xc{itil} strongly helps companies delivering \xc{it} services to improve their processes and to increase their benefits.

\begin{figure}[!t]
	\centering
		\includegraphics[width=1.00\textwidth]{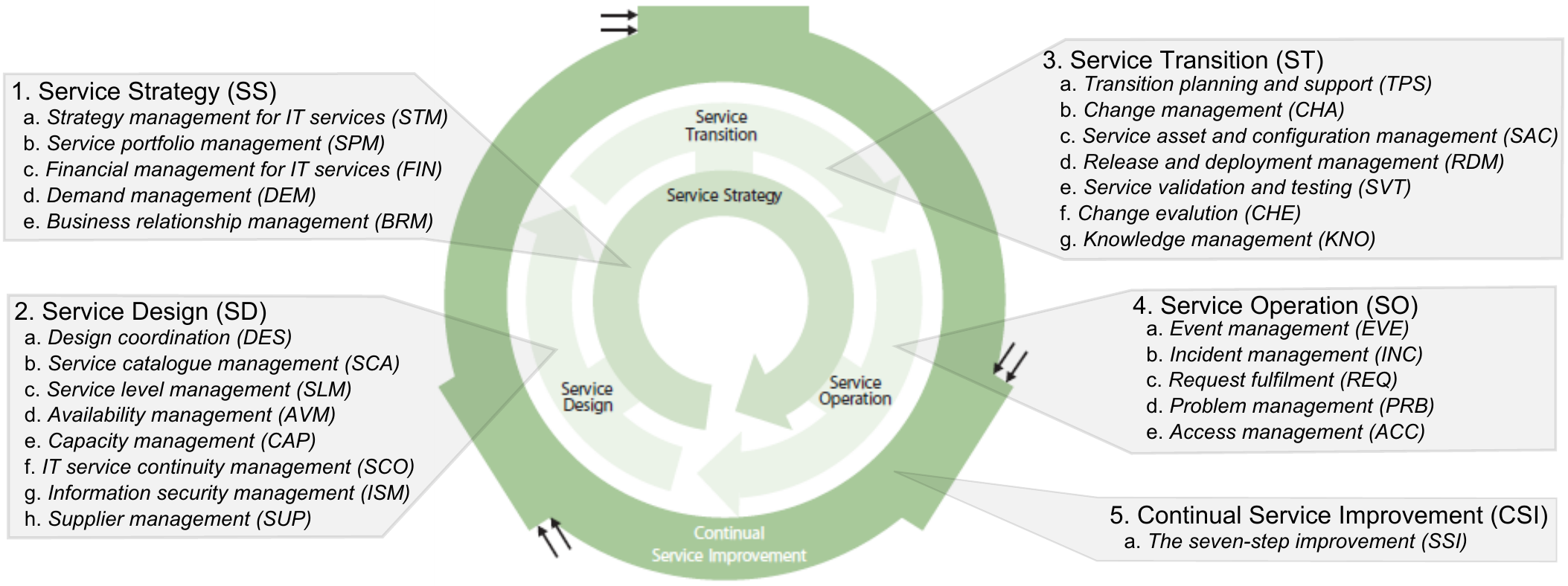}
	\caption{\xc{itil v.}3 life cycle (based on an official illustration of \xc{itil}~\cite{itilSS})}
	\label{fig:ITILv3LC}
\end{figure}

\xc{itil v.}3 is decomposed into five phases which are illustrated in Figure~\ref{fig:ITILv3LC} and described below.

\begin{description}
	\item[Service Strategy] (\xc{ss}): this phase aims at guiding the whole service creation and delivery strategy through an effective management of the life cycle of services. Topics covered in \xc{ss} are mainly the creation and management of the organizational strategy, the analysis and the development of the markets (including internal customers), the feasibility analysis related to the conception of services and the financial management~\cite{itilSS}.
	\item[Service Design] (\xc{sd}): this phase provides the guidance for the conception of services and their future management~\cite{itilSD}. It takes into account the functional as well as the non-functional requirements of the stakeholders. The conception lies in defining how the organization assets will be transformed to bring value to customers through the use of the \xc{it} services provided. This phase also defines how to maintain and increase the value brought to customers and users when service improvement plans are carried out.
	\item[Service Transition] (\xc{st}): this phase explains how to organize the implementation of the designed service solutions and how to manage the changes applied in existing services~\cite{itilST}. These changes are mainly achieved to maintain a service or to increase the value provided to its users.
	\item[Service Operation] (\xc{so}): this phase focuses on the delivery and support of \xc{it} services~\cite{itilSO}. It explains how to manage the service requests and accesses, the events, the incidents and the problems.
	\item[Continual Service Improvement] (\xc{csi}): this phase aims at maintaining the value delivered by \xc{it} services to its users and customers through the measurement and the analysis of the service solutions and the processes followed~\cite{itilCSI}. The output provided are proposals of improvements regarding the quality of the services and the processes followed in the organization.
\end{description}

Each of the \xc{itil} phases is composed of processes. An \xc{itil} process is defined as ``a structured set of activities designed to accomplish a specific objective. A process takes one or more defined inputs and turns them into defined outputs. [...] A process may define policies, standards, guidelines, activities and work instructions if they are needed.''~\cite{itilSS}. For instance, the \xc{ss} phase is composed of five processes: Strategy management for \xc{it} services (1.a \xc{stm}), Service portfolio management (1.b \xc{spm}), Financial management for \xc{it} services (1.c \xc{fin}), Demand management (1.d \xc{dem}) and Business relationship management (1.e \xc{brm}). All processes are named in the illustration showed in Figure~\ref{fig:ITILv3LC}.

\section{A Brief Comparison Between the ITIL v.$3$ Life Cycle and an SoS Implementation Methodology\label{currentMethodology}}
\subsection{Presentation of a Reference Service-Oriented Development Methodology}
\begin{figure}[!t]
	\centering
		\includegraphics[width=0.55\textwidth]{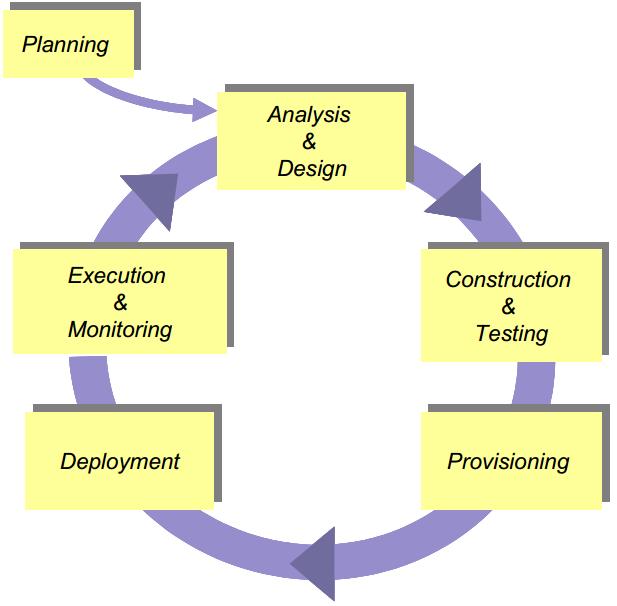}
	\caption{Phases of the service-oriented development methodology~\cite{DBLP:journals/ijwet/PapazoglouH06}}
	\label{fig:papazoglouSOADevelopment}
\end{figure}

In~\cite[chap.~15]{livrePapazoglouWS}, Papazoglou argues that ``there is a clear need for \xc{soa} design methods that allow an organization to avoid the pitfalls of deploying an uncontrolled maze of services and provide a solid foundation for service enablement in an orderly fashion so that Web services can be efficiently used in \xc{soa}-based business applications''. He states that the conventional development methodologies do not tackle the distinctive features of an \xc{soa} such as the composition or the reuse of services. In response, he proposes a service-oriented methodology to implement a service~\cite{DBLP:journals/ijwet/PapazoglouH06}. The phases of this methodology are briefly described hereafter and depicted in Figure~\ref{fig:papazoglouSOADevelopment}.

\begin{enumerate}
	\item \xb{The planning}: during this phase, an analysis of the feasibility and of the scope of the service provisioning solution inside the company environment is achieved. It includes the requirements engineering, the analysis of the current technological situation of the company and the financial estimation of the costs and benefits of the project. The global \xc{ict} vision of the company should also be clear after this phase. The main output is a plan containing the tasks and the description of the deliverables.
	\item \xb{The analysis}: during this phase, the business processes, which will be supported by the future services, must be identified and clearly scoped. Business gaps lead to the modeling or modification of business processes. Moreover, the services must be identified and the ones used in the targeted business processes are known as well as their technical description. The missing functionalities in the \xc{is} are flagged in order to design the corresponding services needed.
	\item \xb{The service design}: during this phase, both the production and execution, as well as the use and composition of the services are designed. The common service concerns such as the reusability, the level of granularity and the composability must be taken into account. The service specifications are composed of three elements: (i) the structural specifications (i.e., the definition of the data types, the exchanged messages and the services interfaces in, for instance, a \xc{wsdl} document), (ii) the behavioral specifications (i.e., the description of the service semantic and of the effects and side effects of the service use, which sometimes lead to the definition of the constraints related to the service use) and (iii) the policy specifications (i.e., the description of the service constraints and its use policy). A clear mapping between the future services and the business process is needed in order to know how they will be composed and which business processes they will support.
	\item \xb{The service construction}: the services are implemented, both at the providers and consumers sides. At the providers side, the service can be created from scratch, composed on the basis of the other existing services or transformed from an existing application into the specified service. At the consumers side, the application component able to communicate with the new service is implemented.
	\item \xb{The service test}: the main objective of this phase is to achieve some dynamic tests on the new service in order to compare its working to its expected behavior. This dynamic tests are mainly the functional, stress, assembly and interface tests. Of course, other kind of tests can also be conducted, such as security, upgrade or compatibility tests.
	\item \xb{The service provisioning}: the main objective of this phase is to support the activities that enable the customers to use the services. This implies both technical aspects (e.g., service auditing and metering) and business aspects (e.g., service governance and billing).
	\item \xb{The service deployment}: during this phase, the new business processes and the underlying (composed) services are rolled out. The most important point is to separate the service building concerns and the deployment concerns.
	\item \xb{The service execution}: this phase contains the activities carried out by the service providers to support the service invocations. This includes dynamic and static service binding.
	\item \xb{The service monitoring}: this last phase aims at measuring and controlling the service quality and its business performance, both at the providers side and the customers side. This step includes the management of the \xc{sla}'s (Service Level Agreement).
\end{enumerate}

Other \xc{soa} development methodologies have been proposed in the literature. A recent publication argues that most of the \xc{soa} methodologies are divided into the six following phases~\cite{DBLP:conf/balt/Svanidzaite12}: Service-oriented analysis, Service-oriented design, Service development/construction, Service testing, Service deployment/transition, Service administration/management. These phases are very close to the ones proposed by Papazoglou in his own methodology. We decided to work with his methodology because it is well-rated compared to other initiatives~\cite{DBLP:journals/soca/GuL11}. Moreover, the main other \xc{sose} methodologies have all at least one significant weakness, e.g., the \xc{soma} methodology proposed by \xc{ibm} is proprietary and very prescriptive~\cite{DBLP:journals/ibmsj/ArsanjaniGAAGH08}, the \xc{soaf} methodology is incomplete because it does not cover all the six phases mentioned above~\cite{DBLP:conf/IEEEscc/ErradiAK06} and the methodology proposed by Erl does not tackle the project issue and is mainly oriented on technical considerations~\cite{erl2007soaPrinciples}. The reader can refer to~\cite{DBLP:conf/balt/Svanidzaite12} and~\cite{ramollari2007survey} for a more complete comparison between \xc{soa} implementation methodologies.

The chosen service-oriented development methodology is mainly composed of guidelines to specify, build and compose the services. One of the primary objectives is to support dynamic business processes with an \xc{is}. However, current companies also require a global view on the management of their services, i.e., they want to adopt an Information Technology Service Management (\xc{itsm}) framework~\cite{DBLP:journals/cacm/GalupDQC09} such as \xc{itil}. While the service-oriented implementation methodology is more about the implementation of services, the reusability and the composability, \xc{itil} focuses on the organizational processes to follow in order to deliver value to the customers and users of services by applying a proper service delivery and support. Even tough the service-oriented paradigm and \xc{itil} seem to be complementary in an organization, the combined use of \xc{itil} and the service-oriented development methodology raises several problems and issues. Some of them are proposed hereafter (\S\ref{CurSitIssue}).

\subsection{Significant Issues Observed in the Current Situation\label{CurSitIssue}}
First of all, the notion of service is defined and comprehended very differently. In \xc{itil} a service, called an \xc{it} service, is defined as the ``means of delivering value to customers by facilitating outcomes customers want to achieve without the ownership of specific costs and risks''~\cite{itilSS}. In the literature concerning the service-oriented paradigm, the definition of the service concept varies somewhat. However, the definition proposed by Papazoglou is often used: a service ``is a self-describing, self-contained software module available via a network [...] which completes tasks, solves problems, or conducts transactions on behalf of a user or application. [...] Services constitute a distributed computer infrastructure made up of many different interacting application modules trying to communicate over private or public networks [...] to virtually form a single logical system''~\cite[chap. 1]{livrePapazoglouWS}. They are several other concepts in \xc{itil} and in the service-oriented paradigm which have similar meaning or syntax such as the notion of \xc{sla}. This will be discussed and solved in~\S\ref{AlignmentSoSprovisioningModel}.
\\A second observation concerns the lack of understanding between the management of \xc{it} organizations and the technical teams operating the hardware and software infrastructure. Moreover, some concepts and terms are understood differently, which causes some confusion between stakeholders. From one side, \xc{itil} helps to establish, structure and improve the management of organizations at the different layers of responsibilities. From the other side, the service-oriented paradigm mainly focuses on the \xc{is} structure and its technical management. However, both of these two organizational layers recommend to adopt a service-orientation. How to combine these two visions of the service-orientation in organizations is not clear~\cite{DBLP:journals/bise/FischbachPA13}. As an example, we can mention the notion of service registry in the service-oriented paradigm and the notion of service portfolio and service catalogue in \xc{itil}. The questions raised are, e.g., related to the similarities and differences between the notions or to the way they can be combined and used as a whole in the organization. Similar issues result from the comparison of the service design package in \xc{itil} and the service description in the service-oriented paradigm.
\\Thirdly, the life cycle of a service is differently structured. The phase Continual Service Improvement (\xc{csi}) of \xc{itil} organizes the improvement of the service solutions and processes. One of its purposes is to align the service proposed with changing and new business needs expressed by the users and customers. In a \xc{s}o\xc{s}, the service monitoring phase focuses on the measurement of the quality of service characteristics~\cite{DBLP:journals/ijwet/PapazoglouH06}. Nothing is said about the new and changing business needs. Therefore, using \xc{itil} to manage the services of a \xc{s}o\xc{s} should help to improve these services.
\\A last observation is the lack of links between the \xc{s}o\xc{s} implementation methodology chosen and a global service management strategy, i.e., an \xc{itsm} framework. Note this observation is also achieved in the main other \xc{sose} methodologies mentioned above. This work contributes to clarify and answer to most of these issues.

\section{The Spiral Model Revisited for the Service-Oriented Paradigm\label{SpiralModelRevisited}}
In order to make clear the links between \xc{itil} and the phases of the service-oriented development methodology, we expand and detail it. We call it the \xc{s}o\xc{s} provisioning model. It is illustrated in Figure~\ref{fig:SOSprovisioningLifeCycle} and describe in~\S\ref{AlignmentSoSprovisioningModel}. The structure in five parts that we use comes from the Spiral Model~\cite{DBLP:journals/computer/Boehm88,DBLP:journals/computer/BoehmEKPSM98} which respects the iterative and incremental process used by Papazoglou to represent the service-oriented design and development methodology~\cite{DBLP:journals/ijwet/PapazoglouH06}. The main objectives of this structure are to answer two fundamental issues when creating an \xc{is}: (i) \zi{What are the objectives and the output of the current stage?} and (ii) \zi{After this stage, what should we do?} Note, once the \xc{s}o\xc{s} provisioning model aligned with \xc{itil v.}3, the way to perform each step will be defined thanks to the \xc{itil} process description.
\\The Spiral Model used in the scope of this work is close to the initial model proposed by Boehm~\cite{DBLP:journals/computer/Boehm88} and its revised version~\cite{DBLP:journals/computer/BoehmEKPSM98}. Nevertheless, we lightly adapt it for the service-oriented paradigm --note the flexibility was one of the main strengths of the Spiral Model and its evolutions called the WinWin Spiral Model~\cite{DBLP:journals/computer/BoehmEKPSM98}. Its structure is composed of five parts numbered with Roman numerals in Figure~\ref{fig:SOSprovisioningLifeCycle}.

\begin{enumerate}[label=\Roman*]
	\item \xb{People identification \& communication}: The steps included in this part focus on the stakeholders. The latter are first identified in order to establish an effective communication between them and the development team in order to keep them informed about the progress of the projects. These steps will also help to manage the organizational changes due to the modifications carried out.
	\item \xb{Determining objectives, alternatives \& constraints}: The steps contained in this part serve to establish the vision and the direction of the project. They help to determine the main objectives of the project as well as its scope, its limits and its constraints. They also help to solve design conflicts after their communication to the stakeholders.
	\item \xb{Risks analysis}: The steps of this part focus on the risk management. Once the vision determined and the choices made, their underlying risks are identified and analysed. A good risk management will help the correct execution of the next steps in which the components of the \xc{s}o\xc{s} is effectively designed, built and deployed.
	\item \textbf{\textsc{s}o\textsc{s} conception \& development}: The steps of this part enable to define the \xc{s}o\xc{s} elements as well as to develop its services --from a business and a technical point of view-- and its components.
	\item \xb{Solution evaluation \& verification}: This fifth and last part includes the steps related to the evaluation of the output achieved during each cycle of the Spiral Model. The end of this part consists of the creation of the planning of the next cycle in which the tasks are placed on a timeline, the resources are identified and then allocated to the tasks.
\end{enumerate}
The main difference with the published Spiral Model is the part numbered I. Indeed, it was not mentioned in the first version of the Spiral Model~\cite{DBLP:journals/computer/Boehm88}. In the WinWin Spiral Model, the importance of the communication with the stakeholders is further considered~\cite{DBLP:journals/computer/BoehmEKPSM98} but the model is quite complicated with seven different stages. In the scope of this paper, we aggregate these three first stages into one in order to ease the understanding and the use of the proposed \xc{s}o\xc{s} provisioning model.

\section{The Alignment of the SoS Provisioning Model with the ITIL v.$3$ Life Cycle\label{AlignmentSoSprovisioningModel}}

\begin{figure}[!t]
	\centering
		\includegraphics[width=1.15\textwidth, angle=90]{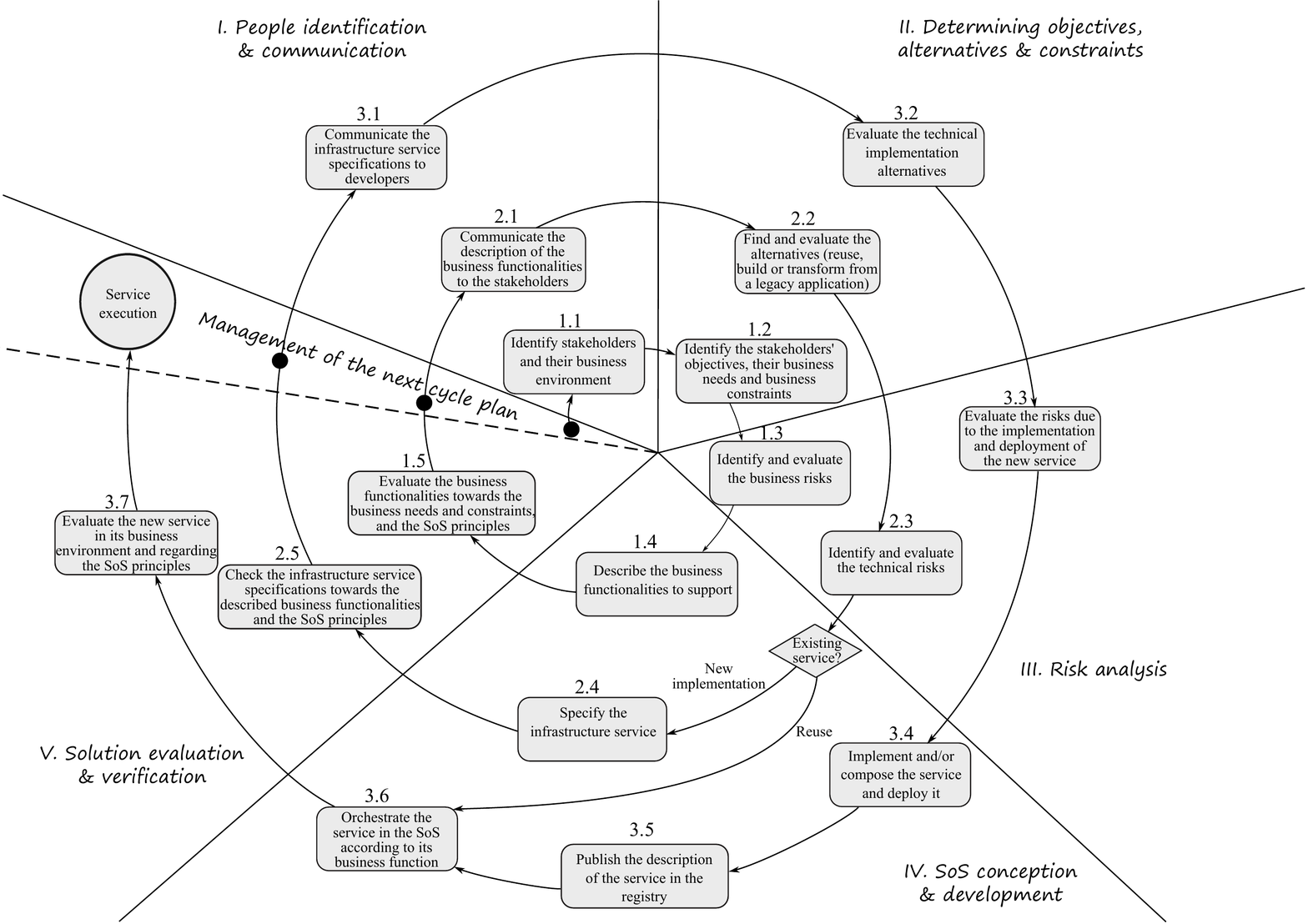}
	\caption{Illustration of the \xc{s}o\xc{s} provisioning model}
	\label{fig:SOSprovisioningLifeCycle}
\end{figure}

The \xc{s}o\xc{s} provisioning\footnote{Note that the concept of \xc{s}o\xc{s} provisioning should be distinguished from the concept of service provisioning.} is the process leading to the analysis, the design, the creation and the composition of services (or the reuse of existing software components) to support, directly or indirectly, some business activities. Figure~\ref{fig:SOSprovisioningLifeCycle} depicts the life cycle of the \xc{s}o\xc{s} provisioning model. This model should be covered for each required business service --a business service is a logical part of an \xc{s}o\xc{s} aligned on an activity of a business process which represents some required business functionalities~\cite{DBLP:journals/ijcis/PapazoglouTDL08}.
\\In the scope of our work, another important concept is the notion of infrastructure service. This is defined as a container along with the service management and monitoring infrastructure that encapsulate computational resources~\cite{DBLP:journals/ijcis/PapazoglouTDL08}. Combined, these infrastructure services can provide the business functionalities needed to become a business service.
\\As a reminder, the notion of service in \xc{itil}, called an \xc{it} service, is defined as the ``means of delivering value to customers by facilitating outcomes customers want to achieve without the ownership of specific costs and risks''~\cite{itilSS}. Compared to the service-oriented paradigm, this definition of the term ``service'' is close to the notion of business service defined above. Indeed, supporting the functionalities of business processes should provide value to the stakeholders by facilitating their business outcomes. Moreover, the use of the service provided by a \xc{s}o\xc{s} allows the transfer of some costs and risks from the stakeholders to the technical staff maintaining the \xc{s}o\xc{s}. In the definition of an \xc{it} service, the term ``means'' refers to, i.a., the infrastructure services supporting the business service since the notion of infrastructure service should be understood as a configuration item in \xc{itil}\footnote{In \xc{itil}, a configuration item is ``any component or other service asset that need to be managed in order to deliver an \xc{it}~service''~\cite{itilST}.}.
\\We recommend to only use the notion of \xc{it} service and infrastructure service given that the notion of business service is redundant with the notion of \xc{it} service. Note that these two perspectives over the notion of service --i.e., the \xc{it} service and the infrastructure service-- should be designed with the appropriate level of granularity during the provisioning of a \xc{s}o\xc{s}. The \xc{it} management of the \xc{s}o\xc{s}, which is made possible with the \xc{itil} framework, helps to achieve this objective.

The following three sections (\S\ref{SoSProvCycle1} to~\S\ref{SoSProvCycle3}) respectively detail the first, the second and the third cycle of the \xc{s}o\xc{s} provisioning model illustrated in Figure~\ref{fig:SOSprovisioningLifeCycle} along with, for each step, the links with the \xc{itil v.}3 processes.

\subsection{Description and Alignment of the First Cycle of the SoS Provisioning Model\label{SoSProvCycle1}}
The first cycle of the \xc{s}o\xc{s} provisioning model focuses on the analysis of the business environment --i.e., the analysis of the stakeholders, their requirements, the business risks and the business constraints-- that the future \xc{it} service will support.

Before the first step numbered 1.1, the first cycle is planned (see ``Management of the next cycle plan'' in Figure~\ref{fig:SOSprovisioningLifeCycle}). The objective is to organize all the steps of the first cycle by planning and structuring the tasks, allocating the resources needed and monitoring the achievement of each step. Regarding the alignment with \xc{itil v.}3, this planning work has to be performed in accordance with the company strategy (defined thanks to the Strategy Management Process --this is the process 1.a \xc{stm} in Figure~\ref{fig:SOSprovisioningLifeCycle}). Then, the Service Portfolio Management Process (1.b \xc{spm}) and, more precisely, the service pipeline activities take the responsibility to organize the tasks leading to the description of the business functionalities. These activities are detailed in the rest of this section.

\titleSoS{1.1 Identify stakeholders and their business environment}: After the identification of the stakeholders, this step aims at having a first contact with them. One key aspect to understand is the context of the business stakeholders in order to know what is their job and how they work.
\\This step is aligned with 1.d \zi{Demand management} and 1.e \zi{Business relationship management}.

\titleITILprocess{1.d \xc{dem}}{The \xc{dem} process has to define the customer and user profiles. It will help during the analysis of their different needs and the business activities that the future services will support.
\newline In the list below, the main inputs (\xc{i}) and the outputs (\xc{o}) of the current \xc{s}o\xc{s} step are cited along with their respective correspondents of the \xc{brm} process.}

\tabIO{	
\begin{tabular}{l|m{\tabColUn} @{ $\leftrightarrow$ } m{\tabColDeux}}
				 & \centering 1.1 Identify stakeholders and their business environment & \centering 1.d \xc{dem} \tabularnewline\hline
	\multirow{2}{*}{\xc{i}}  & List of the services already used by the requesters & Customer portfolio \\
													 & Documentation related to the demand for creating a new service & Initiatives to create or change a service \\\hline
	\multirow{2}{*}{\xc{o}}  & Definition of the user role and their main activities & User profiles \\
													 & Definition of the business processes of the stakeholders and future users & Patterns of business activity\\
\end{tabular}}

\titleITILprocess{1.e \xc{brm}}{The main activities of the \xc{brm} process focus on the identification of potential customers (internal and external) and the stakeholders, and their business context. This process also starts the gathering of the new needs of existing service users (mainly managed at step 1.2).}

\tabIO{
\begin{tabular}{l|m{\tabColUn} @{ $\leftrightarrow$ } m{\tabColDeux}}
	 & \centering 1.1 Identify stakeholders and their business environment & \centering 1.e \xc{brm} \tabularnewline\hline		 
	\multirow{3}{*}{\xc{i}} & Initial demand to create a service & Customer requests \\ 
													& Analysis of the user habits and the related business processes (as-is) & Patterns of business activity \\ 
													& First meeting with the stakeholders and identification of their roles in the organization & User profiles \\\hline
	\multirow{2}{*}{\xc{o}} & Stakeholder register & Stakeholder definitions \\
													& Strengths and weaknesses in the business processes (as-is) & Defined business outcomes \\
\end{tabular}}

\titleSoS{1.2 Identify the stakeholders' objectives, their business needs and business constraints}: This step should help to clarify the business environment of the stakeholders as well as their requirements. The identification and analysis of the business constraints should help to design a feasible service solution.
\\This step is aligned with 1.a \zi{Strategy management for \xc{it} services} and 1.e \zi{Business relationship management}.

\titleITILprocess{1.a \xc{stm}}{The \xc{stm} process allows the service provider to establish its strategy. \xc{itil} first focuses on the business value provided by services to its users. Therefore, the strategy of service providers is mostly focused on business analysis and its position on the targeted markets. This is why the stakeholders' objectives, their business needs and the business constraints will undergo an initial check to be sure that they are in line with the strategy of the service provider and its implementation. A more complete verification is achieved in step 1.5.}
\tabIO{	
\begin{tabular}{l|m{\tabColUn} @{ $\leftrightarrow$ } m{\tabColDeux}}
				 & \centering 1.2 Identify the stakeholders' objectives, their business needs and business constraints & \centering 1.a \xc{stm} \tabularnewline\hline
	\multirow{3}{*}{\xc{i}} & Current strategic business objectives & Existing plans \\
													& Meetings with identified stakeholders & Customer interviews \\
													& Services proposed in the \xc{s}o\xc{s} & Service portfolio \\\hline
	\multirow{2}{*}{\xc{o}} & Main stakeholders business needs & Service strategy\\
													& Prioritized selected business needs & Mission and vision statements \\
\end{tabular}}
	
\titleITILprocess{1.e \xc{brm}}{The activities of the \xc{brm} process begun during the previous step continue in order to establish the main guidelines for the service implementation.}
\tabIO{
\begin{tabular}{l|m{6.5cm} @{ \ $\leftrightarrow$ \ } m{\tabColDeux}}
	 & \centering 1.2 Identify the stakeholders' objectives, their business needs and business constraints & \centering 1.e \xc{brm} \tabularnewline\hline		 
	\multirow{2}{*}{\xc{i}} & Meetings and interviews of the stakeholders, and the business processes (as-is) & Customer requirements \\ 
													& Meetings and interviews of the stakeholders & Customer strategy \\\hline
	\multirow{1}{*}{\xc{o}} & Guidelines for the analysis, design creation of the service & Service requirements for strategy, design and transition\\ 
\end{tabular}}

\titleSoS{1.3 Identify and evaluate the business risks}: This step focuses on the analysis of the risks due to the future use of a \xc{it} service and its possible consequences on the business processes.
\\This step is aligned with 1.b \zi{Service portfolio management} and 1.c \zi{Financial management}. 

\titleITILprocess{1.b \xc{spm}}{The risk analysis related to the use or the implementation of a new \xc{it} service is carried out by the \xc{spm} process. It has to ensure that the requirements of the stakeholders are clearly defined as well as their underlying business risks.}

\tabIO{	
\begin{tabular}{l|m{\tabColUn} @{ $\leftrightarrow$ } m{\tabColDeux}}
				 & \centering 1.3 Identify and evaluate the business risks & \centering 1.b \xc{spm} \tabularnewline\hline	 
	\multirow{2}{*}{\xc{i}} & Program of service implementation projects & Strategy plans\\
													& Prioritization of the selected needs (along with the identified conflicts between stakeholders) & Requests and complaints of stakeholders \\\hline
	\multirow{3}{*}{\xc{o}} & Business agreement for implementing the new service & Service charter \\
													& Addition of the new service in the list of under implementation services & Up-to-date service portfolio \\
													& List of identified business risks to manage during the service implementation & Identified strategic risks \\		
\end{tabular}}

\titleITILprocess{1.c \xc{fin}}{The financial considerations are taken into account by the \xc{fin} process. More specifically, it has to ensure that the price of the service can be paid by its consumers in comparison to the engaged and expected costs. It also arranges the pricing and billing, which helps to manage and control the business risks.}

\tabIO{
\begin{tabular}{l|m{\tabColUn} @{ $\leftrightarrow$ } m{\tabColDeux}}
				 & \centering 1.3 Identify and evaluate the business risks  & \centering 1.c \xc{fin} \tabularnewline\hline 
	\multirow{2}{*}{\xc{i}} & Local regulation and financial governance of the company & Policies, standards and practices related to the financial aspects \\
													& Available budget for the service implementation and a first estimation of the costs & Data sources related to the financial health \\\hline
	\multirow{1}{*}{\xc{o}} & Analysis of the opportunity cost related to the design and implement the service & Service valuation (cost vs. business value) \\
\end{tabular}}

\titleSoS{1.4 Describe the business functionalities to support}: Based on the objectives expressed by the stakeholders, the business constraints and the business risk analysis, this step should lead to the business design of the future service. In other words, the functionalities of the future \xc{it} service are identified and specified.
\\This step is aligned with 1.b \zi{Service portfolio management} and 1.d \zi{Demand management}.

\titleITILprocess{1.b \xc{spm}}{The service charter document contains all the business details of the new \xc{it} service.}

\tabIO{
\begin{tabular}{l|m{\tabColUn} @{ $\leftrightarrow$ } m{\tabColDeux}}
				 & \centering 1.4 Describe the business functionalities to support & \centering 1.b \xc{spm} \tabularnewline\hline
										 
	\multirow{1}{*}{\xc{i}} & Prioritization of the selected needs and optional requirements & Requests, suggestions and complaints of stakeholders \\\hline
	\multirow{2}{*}{\xc{o}} & Business specifications of the service & Service charter \\
													& Possible modifications of the business processes due to the implementation of the new service & Change proposals \\
\end{tabular}}

\titleITILprocess{1.d \xc{dem}}{The \xc{dem} processes end the work started at step 1.1. Under the identified business constraints, the requirements are specified, the conflicts are removed and the future targeted quality levels are defined. Later, these pieces of information will be used to create a document called the \xc{slr} (\xc{s}ervice \xc{l}evel \xc{r}equirements) in \xc{itil}.}

\tabIO{	
\begin{tabular}{l|m{\tabColUn} @{ $\leftrightarrow$ } m{\tabColDeux}}
				 & \centering 1.4 Describe the business functionalities to support & \centering 1.d \xc{dem} \tabularnewline\hline		 
	\multirow{2}{*}{\xc{i}} & Analysis of the workload of existing services & Service and customer portfolio \\
													& Requirements expressed by the stakeholders & Service model needs \\\hline
	\multirow{2}{*}{\xc{o}} & Business specifications of the service and its possible options & Documentation of service package options \\
													& Targeted \xc{q}o\xc{s} and service management policy & Policies for management of demand\\
\end{tabular}}	

\titleSoS{1.5 Evaluate the business functionalities towards the business needs and constraints, and the \xc{s}o\xc{s} principles}: This step ends the first cycle of the \xc{s}o\xc{s} provisioning Model. It should provide an evaluation of the specified \xc{it} service. A comparison with the stakeholders' business needs and with the \xc{s}o\xc{s} principles\footnote{The \xc{s}o\xc{s} principles should be defined when the \xc{s}o\xc{s} has been been implemented.} will help to rate the quality of the specified \xc{it} service and its feasibility.
\\This step is aligned with 1.a \zi{Strategy management for \xc{it} services} and 1.b \zi{Service portfolio management}.
\titleITILprocess{1.a \xc{stm}}{The strategy of the service provider is, i.a., contained in the \xc{s}o\xc{s} principles. The \xc{stm} process achieves a verification of the specified \xc{it} service to make sure that these principles and the service provider strategy are respected.}

\tabIO{	
\begin{tabular}{l|m{\tabColUn} @{ $\leftrightarrow$ } m{\tabColDeux}}
				 & \centering 1.5 Evaluate the business functionalities towards the business needs and constraints, and the \xc{s}o\xc{s} principles & \centering 1.a \xc{stm} \tabularnewline\hline
										 
	\multirow{3}{*}{\xc{i}} & Current strategic business objectives and the \xc{s}o\xc{s} policy & Existing plans \\
													& Analyzed requirements coming from the stakeholders & Customer interviews \\
													& Global analysis of the existing and under development services & Vendor strategies and product roadmaps \\\hline
	\multirow{2}{*}{\xc{o}} & Defined implementation strategy & Tactical plans \\
													& Necessary conditions for the implementation of the service  & Strategic requirements for new services \\
\end{tabular}}	
	
\titleITILprocess{1.b \xc{spm}}{The \xc{spm} process checks if the business needs and constraints are respected by the specified business functionalities. This is the needed agreement for the service charter.}
\tabIO{	
\begin{tabular}{l|m{\tabColUn} @{ $\leftrightarrow$ } m{\tabColDeux}}
				 & \centering 1.5 Evaluate the business functionalities towards the business needs and constraints, and the \xc{s}o\xc{s} principles  & \centering 1.b \xc{spm} \tabularnewline\hline
										 
	\multirow{2}{*}{\xc{i}} & Prioritization of the selected needs and optional requirements along with the feedback given by the stakeholders & Request, suggestions or complaints of stakeholders \\
													& Most part of the analysis and business specification achieved during the step 1 & Project updates for service in the charter stage \\\hline
	\multirow{1}{*}{\xc{o}} & Justified agreement for continuing the \xc{s}o\xc{s} provisioning steps & Authorization for designing and building a new service (service charter) \\
\end{tabular}}

\subsection{Description and Alignment of the Second Cycle of the SoS Provisioning Model\label{SoSProvCycle2}}
The second cycle focuses on the analysis of the technical alternatives that match the business functionalities described and validated during the first cycle. This consists in analyzing the implementation alternatives and the risks of these alternatives, and in specifying the future service.
\\Note, if the business needs can already be satisfied by an existing (composed) service, a part of the second and the third cycle is skipped. Indeed, between the steps $2.3$ and $2.4$, the flow followed depends on the alternative chosen (see Figure~\ref{fig:SOSprovisioningLifeCycle}). In case of service reuse, the process flow leads to the step $3.7$. In all other cases --i.e., implementation of (a part of) the service or modification of a legacy application-- the normal flow must be followed.

First of all, the activities of second cycle are planned and organized (see ``Management of the next cycle plan'' in Figure~\ref{fig:SOSprovisioningLifeCycle}) based on the results obtained from the steps of the first cycle. This means planning and structuring the tasks, allocating the resources needed and monitoring the achievement of each next step. The \xc{itil} process 2.a \zi{Design coordination} is in charge of the organization of the activities related to the service design which leads to the creation of the service design package. These activities are detailed in the rest of this section.

\titleSoS{2.1 Communicate the description of the business functionalities to the stakeholders}: The evaluation performed during the step $1.5$ as well as the specifications of the future service are communicated to the stakeholders, including the \xc{it} management staff.
\\This step is aligned with 1.e \zi{Business relationship management}.

\titleITILprocess{1.e \xc{brm}}{The result of the first cycle of the \xc{s}o\xc{s} provisioning model is communicated to the future service customer and their stakeholders.}
\tabIO{	
\begin{tabular}{l|m{\tabColUn} @{ $\leftrightarrow$ } m{\tabColDeux}}
				 & \centering 2.1 Communicate the description of the business functionalities to the stakeholders & \centering 1.e \xc{brm} \tabularnewline\hline			 
	\multirow{1}{*}{\xc{i}} & Business specifications of the new service and listing of the stakeholders & Project portfolio \\\hline
	\multirow{2}{*}{\xc{o}} & Communication of the main needs and constraints for the next steps (coming for the business service specifications) & Service requirements for strategy, design and transition \\
													& Schedule communication to the stakeholders & Schedules of customer/stakeholder activity \& Schedule of awareness events\\
													
\end{tabular}}	

\titleSoS{2.2 Find and evaluate the alternatives (reuse, build or transform from a legacy application)}: Based on the exchanged information achieved during the previous step, the possible solution(s) will be considered. These alternatives can be of three types: (i) service reuse --an existing service is used; it can be provided by the \xc{s}o\xc{s} of the organization or by an external service provider--, (ii) building of the service --the service functionalities will be built from scratch, and/or existing services will be composed to support the functionalities needed to provide the \xc{it} service--, or (iii) the service functionalities exist in a legacy application --in this case, the software component will be encapsulated by the service standards commonly used in the \xc{s}o\xc{s}.
\\This step is aligned with 2.b \zi{Service catalogue management}.

\titleITILprocess{2.b \xc{sca}}{The business service specifications, which are described in the service charter document, are compared to the existing services available in the service catalogue(s). The objective is to find an existing service or a combination of existing services that cover the business needs expressed in the service charter.}

\tabIO{	
\begin{tabular}{l|m{\tabColUn} @{ $\leftrightarrow$ } m{\tabColDeux}}
				 & \centering 2.2 Find and evaluate the alternatives (reuse, build or transform from a legacy application) & \centering 2.b \xc{sca} \tabularnewline\hline			 
	\multirow{2}{*}{\xc{i}} & Business specifications of the service & Business and financial information, and business requirements \\
													& Result from the service evaluation and customer requirements & Business impact analysis \\\hline
	\multirow{2}{*}{\xc{o}} & Analysis of the implementation alternatives & Documentation of a definition of the service \\
													& Justification and documentation of the choice made among the alternatives & Update to the service portfolio \\
\end{tabular}}

\titleSoS{2.3 Identify and evaluate the technical risks}: This step aims at identifying and evaluating the risks raised by the alternative chosen at the previous step. These risks are related to the existing \xc{is}s, the other ongoing service implementation projects and the other existing services in use. This technical risk analysis completes the business risk analysis achieved during the step 1.3. This step provides relevant information to specify the infrastructure service. If the reuse alternative is chosen, the current availability, capacity, service continuity and information security is compared to the requirements and to the business service specifications.
\\This step is aligned with 2.d \zi{Availability management}, 2.e \zi{Capacity management}, 2.f \zi{\xc{it} service continuity management} and 2.g \zi{Information security management}.

\titleITILprocess{2.d \xc{avm}}{If the chosen alternative is the reuse of an existing service, the availability of this existing service will be assessed according to the new needs identified during the step 1.2. Otherwise, the technical risks related to the availability of the existing services and infrastructure are identified and assessed.}

\tabIO{	
\begin{tabular}{l|m{\tabColUn} @{ $\leftrightarrow$ } m{\tabColDeux}}
				 & \centering 2.3 Identify and evaluate the technical risks & \centering 2.d \xc{avm} \tabularnewline\hline			 
	\multirow{3}{*}{\xc{i}} & Business specifications of the service & Availability-related requirements and business impact information\\
													& \xc{s}o\xc{s} monitoring information & Past performance\\
													& Technical specifications of the existing components and infrastructure & Technology information\\\hline
	\multirow{2}{*}{\xc{o}} & \zi{(In case of service reuse)} Service availability forecasting & Availability plan\\
													& \zi{(Otherwise)} Criteria of service availability and related identified risks & Availability and recovery design criteria\\
\end{tabular}}
\titleITILprocess{2.e \xc{cap}}{Similarly to the \xc{avm} process, the capacity of the existing service reused is evaluated if the reuse alternative is the one chosen. Otherwise, the capacity of the existing infrastructure is evaluated in order to identify and evaluate the technical risks.}

\tabIO{	
\begin{tabular}{l|m{\tabColUn} @{ $\leftrightarrow$ } m{\tabColDeux}}
				 & \centering 2.3 Identify and evaluate the technical risks & \centering 2.e \xc{cap} \tabularnewline\hline			 
	\multirow{2}{*}{\xc{i}} & Business specifications of the service & Capacity-related requirements and business impact information\\
													& \xc{s}o\xc{s} monitoring information and use statistics of the existing services & Workload and performance information  \\\hline
	\multirow{2}{*}{\xc{o}} & \zi{(In case of service reuse)} Service capacity forecasting & Capacity plan and service performance information \\
													& \zi{(Otherwise)} Service capacity criteria and related identified risks & Workload analysis, and thresholds, alerts and events \\
\end{tabular}}

\titleITILprocess{2.f \xc{sco}}{Whatever the alternative chosen, the technical risks (except information security risks) are identified, described and analysed thanks to the \xc{sco} process. The objective is to define or adapt continuity plans in accordance with the type of business processes supported by the specified service and with the future customers and users.}

\tabIO{	
\begin{tabular}{l|m{\tabColUn} @{ $\leftrightarrow$ } m{\tabColDeux}}
				 & \centering 2.3 Identify and evaluate the technical risks & \centering 2.f \xc{sco} \tabularnewline\hline			 
	\multirow{2}{*}{\xc{i}} & Business specifications of the service & Business information and the business continuity strategy\\
													& \zi{(In case of service reuse)} Past information on the service continuity  & \xc{it} service continuity plans and test reports \\\hline
	\multirow{2}{*}{\xc{o}} & Service continuity criteria & \xc{itscm} plans\\
													& Report on technical risks & Risks assessment and reports\\
\end{tabular}}

\titleITILprocess{2.g \xc{ism}}{The security risks related to the alternative chosen are identified and analyzed by the \xc{ism} process.}

\tabIO{	
\begin{tabular}{l|m{\tabColUn} @{ $\leftrightarrow$ } m{\tabColDeux}}
				 & \centering 2.3 Identify and evaluate the technical risks & \centering 2.g \xc{ism} \tabularnewline\hline			 
	\multirow{3}{*}{\xc{i}} & Business specifications of the service & Business information\\
													& \xc{s}o\xc{s} security policy and principles & Security governance \\
													& Security information from the \xc{s}o\xc{s} monitoring & Details of all security events and breaches\\\hline
	\multirow{2}{*}{\xc{o}} & \zi{(In case of service reuse)} Security report on the existing service & Security risks assessment and reports \\
													& Service security criteria  & Set of security criteria and controls \\
\end{tabular}}

\titleSoS{2.4 Specify the infrastructure service}: Analysts should specify the \xc{it} service functionality(ies) in order to implement the corresponding infrastructure service during the subsequent steps.
\\This step is aligned with 2.d \zi{Availability management}, 2.e \zi{Capacity management}, 2.f \zi{\xc{it} service continuity management}, 2.g \zi{Information security management} and 2.h \zi{Supplier management}.

\titleITILprocess{2.d \xc{avm}}{The availability needed --it is defined in the specified business service, and in the availability plan and criteria-- is implemented in the infrastructure of the future service. If required, redundancy solutions are also integrated into the specifications. Note these specifications of the infrastructure service are summed up in the service design package, which is the basis for the service transition in \xc{itil}.}

\tabIO{	
\begin{tabular}{l|m{\tabColUn} @{ $\leftrightarrow$ } m{\tabColDeux}}
				 & \centering 2.4 Specify the infrastructure service & \centering 2.d \xc{avm} \tabularnewline\hline			 
	\multirow{2}{*}{\xc{i}} & Business specifications of the service & Business information\footnotemark~and the availability requirements \\
													& Specification of the existing \xc{s}o\xc{s} components and technology knowledge of the service designers & Component information and technology information \\\hline
	\multirow{2}{*}{\xc{o}} & Targeted \xc{s}o\xc{s} related to the availability and measurement tools & Service availability, reliability and maintainability plan \\
													& Technical specifications of the service related to its availability & Details of the availability techniques and measures \\
\end{tabular}}
\footnotetext{Note the business information includes the financial information.} 

\titleITILprocess{2.e \xc{cap}}{The needed configuration items and enabling services provided by internal technical teams are determined considering the needs documented in the business service specifications. This is integrated into the specifications of the infrastructure service.}

\tabIO{	
\begin{tabular}{l|m{\tabColUn} @{ $\leftrightarrow$ } m{\tabColDeux}}
				 & \centering 2.4 Specify the infrastructure service & \centering 2.e \xc{cap} \tabularnewline\hline			 
	\multirow{3}{*}{\xc{i}} &  Business specifications of the service & Business information and the capacity requirements \\
													&  Specification of the existing \xc{s}o\xc{s} components and technology knowledge of the service designers & Component performance and capacity information \\
													& \xc{s}o\xc{s} monitoring information & Workload information \\\hline
	\multirow{2}{*}{\xc{o}} & Targeted \xc{s}o\xc{s} related to the capacity and measurement tools & Capacity plan \\
													& Technical specifications of the service related to its capacity & Forecasts, ad hoc capacity, and thresholds, alerts and events\\
\end{tabular}}

\titleITILprocess{2.f \xc{sco}}{The continuity plans, defined during the step 2.3, along with their existing workarounds, are integrated into the specifications of the infrastructure service. The measures defined following the analysis achieved at the previous step are also integrated into the service specifications.}

\tabIO{	
\begin{tabular}{l|m{\tabColUn} @{ $\leftrightarrow$ } m{\tabColDeux}}
				 & \centering 2.4 Specify the infrastructure service & \centering 2.f \xc{sco} \tabularnewline\hline			 
	\multirow{3}{*}{\xc{i}} & Business specifications of the service & Business information, and the service continuity requirements and strategy \\
													& Specification of the existing \xc{s}o\xc{s} components and technology knowledge of the service designers & Component information and service historical reports \\
													& Technical specifications of the service related to its availability and capacity & Capacity and availability management information \\\hline
	\multirow{2}{*}{\xc{o}} & Technical specifications of the service related to its continuity plans & \xc{itscm} plans including crisis management plans, emergency responses and disaster recovery plans \\
													& Description of the service continuity tests & \xc{itscm} test scenarios \\
\end{tabular}}

\titleITILprocess{2.g \xc{ism}}{The information security criteria defined at the step 2.3 are integrated into the specifications of the infrastructure service.}

\tabIO{	
\begin{tabular}{l|m{\tabColUn} @{ $\leftrightarrow$ } m{\tabColDeux}}
				 & \centering 2.4 Specify the infrastructure service & \centering 2.g \xc{ism} \tabularnewline\hline			 
	\multirow{3}{*}{\xc{i}} & Business specifications of the service & Business information and the information security requirements \\
													& \xc{s}o\xc{s} security principles, security risks & Security governance and risks assessment reports\\
													& \xc{IT} component specifications of partners and suppliers & Details of partner and supplier access \\\hline
	\multirow{3}{*}{\xc{o}} & Service security rules for users and stakeholders & Information security management policy \\
													& Specific service security rules for partners and suppliers & Policies and processes for managing partners and suppliers\\
													& Security specifications of the new service & Set of security classifications and classified informations assets \\
\end{tabular}}

\titleITILprocess{2.h \xc{sup}}{If the configuration items or the enabling services will be delivered by external providers, the \xc{sup} process is used to communicate with them, to compare their offers and to establish the contracts called \xc{uc}\footnote{\xc{uc} stands for Underpinning Contract.} in \xc{itil}.}

\tabIO{	
\begin{tabular}{l|m{\tabColUn} @{ $\leftrightarrow$ } m{\tabColDeux}}
				 & \centering 2.4 Specify the infrastructure service & \centering 2.h \xc{sup} \tabularnewline\hline			 
	\multirow{3}{*}{\xc{i}} & Business specifications of the service & Business information \\
													& Existing supplier lists and existing supplier performance reports & Existing contracts and agreements with suppliers \\
													& Reports coming from the monitoring of \xc{it} component provided by the suppliers & Supplier performance issues and report \\\hline
	\multirow{2}{*}{\xc{o}} & Amendments and new contracts  & Supplier contracts \\
													& Specifications of the \xc{it} components provided by suppliers & Reports on agreed supplier performance \\
\end{tabular}}

\titleSoS{2.5 Check the infrastructure service specifications towards the described business functionalities and the \xc{s}o\xc{s} principles}: During this step, the infrastructure service specifications should be validated regarding the \xc{it} service description as well as the \xc{s}o\xc{s} principles which ensure that the \xc{it} governance of the \xc{s}o\xc{s} is respected.
\\This step is aligned with 2.c \zi{Service level management}.

\titleITILprocess{2.c \xc{slm}}{The quality characteristics described in the service design package are compared to the \xc{slr} which is based on the requirements expressed by the stakeholders and the specifications of the business service.}

\tabIO{	
\begin{tabular}{l|m{\tabColUn} @{ $\leftrightarrow$ } m{\tabColDeux}}
				 & \centering 2.5 Check the infrastructure service specifications towards the described business functionalities and the \xc{s}o\xc{s} principles & \centering 2.c \xc{slm} \tabularnewline\hline			 
	\multirow{3}{*}{\xc{i}} & Business specifications of the service and requirements of the stakeholders & Business information and requirements \\
													& Service registry and service documentation & Service portfolio and service catalogue\\
													& Infrastructure specifications of the service & Component information\\\hline
	\multirow{2}{*}{\xc{o}} & Service quality description and documentation & Service quality plan \\
													& Service description and service agreement template for the \xc{q}o\xc{s} & Document templates \\
\end{tabular}}

\subsection{Description and Alignment of the Third Cycle of the SoS Provisioning Model\label{SoSProvCycle3}}
The third cycle focuses on the implementation and deployment of the specified infrastructure service, i.e., the evaluation of the technical choices, the risk management related to the implementation and deployment of services, the coding of the service, the publication of its description and its orchestration to support the business activities analysed during the first cycle.
\\First of all, the activities of the third cycle are planned and organized (see ``Management of the next cycle plan'' in Figure~\ref{fig:SOSprovisioningLifeCycle}) based on the achievement of the steps of the second cycle. This means planning and structuring the tasks, allocating the resources needed and monitoring the achievement of each next steps. The \xc{itil} process 3.a \zi{Transition planning and support} is in charge of this work which leads to the implementation or the reuse of the service. This work is detailed in the rest of this section.

As explained in~\S\ref{SoSProvCycle2}, if the business functions needed are already provided by existing service(s) which can be reused, this last cycle proves to be very short. Indeed, it only contains the two last steps of this third cycle, i.e., the orchestration of the corresponding infrastructure service and a final evaluation.

\titleSoS{3.1 Communicate the infrastructure service specifications to developers}: The validated specifications of the infrastructure are communicated to the developers who will implement the designed service and publish its technical description in the registry.
\\This step is aligned with 3.a \zi{Transition planning and support}.

\titleITILprocess{3.a \xc{tps}}{The \xc{tps} process has to organize the service transition phase, including the communication between the members of the development team.}

\tabIO{	
\begin{tabular}{l|m{\tabColUn} @{ $\leftrightarrow$ } m{\tabColDeux}}
				 & \centering 3.1 Communicate the infrastructure service specifications to developers & \centering 3.a \xc{tps} \tabularnewline\hline			 
	\multirow{1}{*}{\xc{i}} & Specifications of the infrastructure service and \xc{q}o\xc{s} criteria & Service design package \\\hline
	\multirow{1}{*}{\xc{o}} & Overview of the implementation activities to complete & Integrated set of service transition plans \\
\end{tabular}}

\titleSoS{3.2 Evaluate the technical implementation alternatives}: The technical choices such as the development environment to use are made after their evaluation and comparison. This step should also take into account the constraints due to the use of legacy software component(s) to build the new infrastructure service.
\\This step is aligned with 3.b \zi{Change management}, 3.c \zi{Service asset and configuration management} and 3.e \zi{Service validation and testing}.

\titleITILprocess{3.b \xc{cha}}{The service design package is transformed into one or several Request(s) For Change (\xc{rfc}). The technical choices can be assessed, the schedule of changes is drawn and the resources allocated. Based on the information about the existing configuration items and the enabling services, which are registered in the \xc{cmdb} (Configuration Management DataBase), the change records are finalized. They are derived from the \xc{rfc} and they contain all the details of the changes to implement.}

\tabIO{
\begin{tabular}{l|m{\tabColUn} @{ $\leftrightarrow$ } m{\tabColDeux}}
				 & \centering 3.2 Evaluate the technical implementation alternatives & \centering 3.b \xc{cha} \tabularnewline\hline			 
	\multirow{3}{*}{\xc{i}} & Specification of each service component to implement (or change) & \xc{rfc} coming from the service design package\\
													& \xc{s}o\xc{s} principles and governance related to the implementation of services & Policy and strategy for change and release\\
													& Ongoing projects related to the \xc{s}o\xc{s} & Current change schedule \\\hline
	\multirow{2}{*}{\xc{o}} & Description of the implementation tasks & Change decisions (\xc{rfc}s status), documents and actions \\
													& Implementation planning and resources allocation & Revised change schedule and \xc{pso}\\
\end{tabular}}

\titleITILprocess{3.c \xc{sac}}{The \xc{sac} process manages the technical information about the service components (existing and to implement). It helps during the evaluation of the alternatives by providing all the information needed.}

\tabIO{	
\begin{tabular}{l|m{\tabColUn} @{ $\leftrightarrow$ } m{\tabColDeux}}
				 & \centering 3.2 Evaluate the technical implementation alternatives & \centering 3.c \xc{sac} \tabularnewline\hline			 
	\multirow{3}{*}{\xc{i}} & Specification of the infrastructure service and the implementation planning & Designs, plans and configurations (service design packages) \\
													& Description of the implementation tasks & \xc{rfc}s \\
													& Technical documentation related to the existing service components & Actual configurations information \\\hline
	\multirow{1}{*}{\xc{o}}	& Technical description of \xc{s}o\xc{s} components  & Information about attributes and relationships of configuration items\\
\end{tabular}}

\titleITILprocess{3.e \xc{svt}}{The \xc{svt} process begins at this step to set up a validation plan of the alternative chosen. From the \xc{itil v.}3 point of view, the most important element for this process is to ensure the respect of the needs expressed by the stakeholders in terms of utility and warranty --i.e., the specifications of the \xc{it} service.}

\tabIO{	
\begin{tabular}{l|m{\tabColUn} @{ $\leftrightarrow$ } m{\tabColDeux}}
				 & \centering 3.2 Evaluate the technical implementation alternatives & \centering 3.e \xc{svt} \tabularnewline\hline			 
	\multirow{3}{*}{\xc{i}} & Specification of the infrastructure service and targeted \xc{q}o\xc{s} levels & Service design package (including all the acceptance criteria) \\
													& Business specifications of the service & Customer requests (from the service charter) \\
													& Description of the implementation tasks & \xc{rfc}s\\\hline
	\multirow{1}{*}{\xc{o}} & Planning and description of the service tests & Configuration baseline of the testing environment \\						
\end{tabular}}

\titleSoS{3.3 Evaluate the risks due to the implementation and deployment of the new service}: This step focuses on the identification and the management of the technical risks raised by the implementation of a new service in the \xc{s}o\xc{s}.
\\This step is aligned with 3.a \zi{Transition planning and support} and 3.f \zi{Change evaluation}.

\titleITILprocess{3.a \xc{tps}}{The supervision of the service transition phase is achieved by the \xc{tps} process. It will mainly focus on the analysis of the risks due to possible interferences between the current project and other activities in progress. These risks have to be identified, analysed and controlled.}

\tabIO{	
\begin{tabular}{l|m{\tabColUn} @{ $\leftrightarrow$ } m{\tabColDeux}}
				 & \centering 3.3 Evaluate the risks due to the implementation and deployment of the new service & \centering 3.a \xc{tps} \tabularnewline\hline			 
	\multirow{2}{*}{\xc{i}} & Planned implementation tasks & Authorized changes (along with their risks described in the \xc{rfc})\\
													& Specification of the infrastructure service & Service design package\\\hline
	\multirow{1}{*}{\xc{o}} & Validated planning after the comparison with other ongoing implementation projects & Integrated set of service transition plans \\
\end{tabular}}

\titleITILprocess{3.f \xc{che}}{The risks contained in the \xc{rfc} must be evaluated by the \xc{che} process in an interim evaluation report. This activity gives all the required information to the \xc{cab}\footnote{The \xc{c}hange \xc{a}dvisory \xc{b}oard (\xc{cab}) is an entity of the \xc{cha} process, which assesses and authorizes the implementation of the \xc{rfc}s.} in order to let them make a decision about the \xc{rfc}s.}

\tabIO{	
\begin{tabular}{l|m{\tabColUn} @{ $\leftrightarrow$ } m{\tabColDeux}}
				 & \centering 3.3 Evaluate the risks due to the implementation and deployment of the new service & \centering 3.f \xc{che} \tabularnewline\hline			 
	\multirow{3}{*}{\xc{i}} & Specification of the infrastructure service & Service design package \\
													& Description of the implementation tasks & \xc{rfc}s \\
													& Meetings with the team of the project (including the business analysts, designers, architects, etc.) & Discussions with stakeholders\\\hline
	\multirow{1}{*}{\xc{o}} & Report on the implementation and deployment risks & Interim evaluation report(s) for change management (``before building'' report)\\
\end{tabular}}

\titleSoS{3.4 Implement and/or compose the service and deploy it}: During this step, the infrastructure service will be implemented according to its specifications. Eventually, the reused service(s) is/are composed with the new service built. Then the new service is deployed on its hosting application server.
\\This step is aligned with 3.b \zi{Change management}, 3.c \zi{Service asset and configuration management}, 3.d \zi{Release and deployment management} and 3.f \zi{Change evaluation}.

\titleITILprocess{3.b \xc{cha}}{The authorized \xc{rfc} are effectively implemented.}

\tabIO{	
\begin{tabular}{l|m{\tabColUn} @{ $\leftrightarrow$ } m{\tabColDeux}}
				 & \centering 3.4 Implement and/or compose the service and deploy it & \centering 3.b \xc{cha} \tabularnewline\hline			 
	\multirow{3}{*}{\xc{i}} & Description of the implementation tasks to complete & Authorized \xc{rfc}s \\
													& Validated planning and resources allocation  & Plan (change, transition, release and test) \\
													& Report on the implementation and deployment risks & Interim evaluation reports \\\hline
	\multirow{2}{*}{\xc{o}} & Implemented service components & Change to the services or infrastructure resulting from authorized changes \\
													& Results of the implementation tasks on the \xc{s}o\xc{s} components & New, changed or disposed configuration items \\
\end{tabular}}

\titleITILprocess{3.c \xc{sac}}{All the changes achieved on configuration items are recorded and documented in the \xc{cmdb}.}

\tabIO{	
\begin{tabular}{l|m{\tabColUn} @{ $\leftrightarrow$ } m{\tabColDeux}}
				 & \centering 3.4 Implement and/or compose the service and deploy it & \centering 3.c \xc{sac} \tabularnewline\hline			 
	\multirow{2}{*}{\xc{i}} & Specification of the infrastructure service & Service design package \\
													& Implementation tasks description and their results & \xc{rfc}s and work orders from change management \\\hline
	\multirow{1}{*}{\xc{o}} & Updated documentation of the modified \xc{s}o\xc{s} components and of the new service & New and updated configuration records \\
\end{tabular}}

\titleITILprocess{3.d \xc{rdm}}{If the service design package includes the use of existing services to satisfy the \xc{it} service described, these services are composed with the newly service(s) created based on the \xc{rfc}. In all cases, the (composed) service is deployed to be used by service consumers.}

\tabIO{	
\begin{tabular}{l|m{\tabColUn} @{ $\leftrightarrow$ } m{\tabColDeux}}
				 & \centering 3.4 Implement and/or compose the service and deploy it & \centering 3.d \xc{rdm} \tabularnewline\hline			 
	\multirow{5}{*}{\xc{i}} & Planned implementation tasks & Authorized changes \\
													& Specification of the infrastructure service & Service design package \\
													& Technical description of the service components to create or to modify & Acquired service components and their documentation \\
													& Specification and documentation of the integration of the service components & Build models and plans\\
													& \xc{s}o\xc{s} principles related to the service building and release & Release policy \\\hline
	\multirow{2}{*}{\xc{o}} & Creation of the release and deployment planning & Release and deployment plan \\
													& Integrated service components & New and changed services \\
\end{tabular}}

\titleITILprocess{3.f \xc{che}}{After the implementation of the service components and their integration, an evaluation report is carried out.}

\tabIO{	
\begin{tabular}{l|m{\tabColUn} @{ $\leftrightarrow$ } m{\tabColDeux}}
				 & \centering 3.4 Implement and/or compose the service and deploy it & \centering 3.f \xc{che} \tabularnewline\hline			 
	\multirow{3}{*}{\xc{i}} & Specification of the infrastructure service and targeted \xc{q}o\xc{s} levels & Service design package \\
													& Description of the implementation tasks and their documented results & \xc{rfc}s along with the detailed change documentation \\
													& Planning and description of the service tests & Previous reports \\\hline
	\multirow{1}{*}{\xc{o}} & Tests and their analysis of the implemented (composed) service & Evaluation report \\
\end{tabular}}

\titleSoS{3.5 Publish the description of the service in the registry}: Functional and non-functional characteristics of the built service are described as well as the communication process details to follow in order to use this new service. These documents are then published on a registry which allows the discovery of the new service.
\\This step is aligned with 2.b \zi{Service catalogue management}, 2.c \zi{Service level management} and 3.c \zi{Service asset and configuration management}.

\titleITILprocess{2.b \xc{sca}}{In the \xc{s}o\xc{s}, service descriptions are published in a service registry. This service registry corresponds to the service catalogue in \xc{itil}.}. One well-know technology to structure service registries is \xc{uddi}~\cite{uddiSpecV3}.

\tabIO{
\begin{tabular}{l|m{\tabColUn} @{ $\leftrightarrow$ } m{\tabColDeux}}
				 & \centering 3.6 Publish the new service in the registry & \centering 2.b \xc{sca} \tabularnewline\hline			 
	\multirow{1}{*}{\xc{i}} & Notification of the release of a new service & Feedback from the \xc{svt} process \\\hline
	\multirow{1}{*}{\xc{o}} & Registration of the service and publication of its description and its \xc{sla} criteria in the service registry & Updates to the service catalogue \\
\end{tabular}}

\titleITILprocess{2.c \xc{slm}}{The targeted service quality are defined (at step 2.4) and documented by the \xc{slm} process. This information is then used to monitor the actual delivered service quality.}

\tabIO{	
\begin{tabular}{l|m{\tabColUn} @{ $\leftrightarrow$ } m{\tabColDeux}}
				 & \centering 3.5 Publish the description of the service in the registry & \centering 2.c \xc{slm} \tabularnewline\hline			 
	\multirow{2}{*}{\xc{i}} & Implementation and interim test results related to the service & Change information \\
													& Technical information on the service components & Configuration items information \\\hline
	\multirow{1}{*}{\xc{o}} & Modification and validation of the existing \xc{sla} documents (see step 2.5) & Document template\\
\end{tabular}}

\titleITILprocess{3.c \xc{sac}}{The newly built service is described thanks to the information about the configuration item used to set up the service. The list of the configuration items used is available in the service catalogue. They are registered in the \xc{cmdb}, which is managed by the \xc{sac} process. Note one technology commonly used to describe the operational characteristics of a service is \xc{wsdl}~\cite{wsdl20-part1}.}

\tabIO{
\begin{tabular}{l|m{\tabColUn} @{ $\leftrightarrow$ } m{\tabColDeux}}
				 & \centering 3.5 Publish the description of the service in the registry & \centering 3.c \xc{sac} \tabularnewline\hline			 
	\multirow{2}{*}{\xc{i}} & Specification of the infrastructure service & Service design package \\
													& Technical description of the service and its hosting information & \xc{rfc}s and configuration information \\\hline
	\multirow{2}{*}{\xc{o}} & Service hosting information & Updated configuration records \\
													& Description of the functional characteristics of the service & Status reports and consolidated configuration information \\
\end{tabular}}

\titleSoS{3.6 Orchestrate the service in the \xc{s}o\xc{s} according to its business function}: This step consists of the orchestration of the new (or reused) service. It has to be integrated in the existing composite application which supports the business environment of the \xc{it} service analysed during the first cycle of the \xc{s}o\xc{s} provisioning. If the composite application does not exist, it should be built. Note this part is not covered in this work. Indeed, it is not directly related to the creation of a service.
\\This step is aligned with 3.b \zi{Change management} and 3.c \zi{Service asset and configuration management}.

\titleITILprocess{3.b \xc{cha}}{The orchestration of the deployed service with other services supporting the same business activities --from the point of view of the service consumers and users-- is achieved. The possible modifications needed in the configuration items are documented in the \xc{rfc} and carried out.}

\tabIO{	
\begin{tabular}{l|m{\tabColUn} @{ $\leftrightarrow$ } m{\tabColDeux}}
				 & \centering 3.6 Orchestrate the service in the \xc{s}o\xc{s} according to its business function & \centering 3.b \xc{cha} \tabularnewline\hline			 
	\multirow{3}{*}{\xc{i}} & Service composition design and its technical description & \xc{rfc}s (related to the association of services) \\
													& Validated planning and resources allocation & Plan (change, transition, release and test) \\
													& Test results applied on the service components and on the running service & Interim evaluation reports \\\hline
	\multirow{1}{*}{\xc{o}} & Usable and orchestrated service & Modifications to the services or infrastructure resulting from authorized changes \\
\end{tabular}}

\titleITILprocess{3.c \xc{sac}}{The modifications in the configuration items are documented in the \xc{cmdb}.}

\tabIO{	
\begin{tabular}{l|m{\tabColUn} @{ $\leftrightarrow$ } m{\tabColDeux}}
				 & \centering 3.6 Orchestrate the service in the \xc{s}o\xc{s} according to its business function & \centering 3.c \xc{sac} \tabularnewline\hline			 
	\multirow{1}{*}{\xc{i}} & Technical information concerning the orchestration of the service & \xc{rfc}s and configuration information \\\hline
	\multirow{2}{*}{\xc{o}} & Update service information due to its orchestration & Updated configuration records \\
													& Relation with other services due to its orchestration & Information about relationships of configuration items \\
\end{tabular}}

\titleSoS{3.7 Evaluate the new service in its business environment and regarding the \xc{s}o\xc{s} principles}: This last step of the \xc{s}o\xc{s} provisioning model focuses on the validation of the modified composite application. This is based on the service built or reused, in comparison with the business processes supported by the \xc{s}o\xc{s} and affected by the executed modification in the system. The \xc{s}o\xc{s} principles should also be considered during the evaluation of the new service.
\\This step is aligned with 3.e \zi{Service validation and testing}.

\titleITILprocess{3.e \xc{svt}}{The validation plan is followed in order to test and validate the newly deployed service in its business environment.}

\tabIO{	
\begin{tabular}{l|m{\tabColUn} @{ $\leftrightarrow$ } m{\tabColDeux}}
				 & \centering 3.7 Evaluate the new service in its business environment and regarding the \xc{s}o\xc{s} principle & \centering 3.e \xc{svt} \tabularnewline\hline
	\multirow{3}{*}{\xc{i}} & Specification of the infrastructure service and of the business service & Service design package, including the acceptance criteria\\
													& Description of the implementation and orchestration tasks and their documented results & \xc{rfc}s \\
													& Report concerning the implementation and deployment risks, on the service components tests and on the integration tests & Previous evaluation reports \\\hline
	\multirow{1}{*}{\xc{o}} & Report on the actual \xc{q}o\xc{s} obtained & Validation results and their analysis \\
\end{tabular}}

\subsection{Concluding Remarks Concerning the SoS Provisioning Model Aligned with ITIL v.$3$\label{AligmentConcludingRmqs}}
\begin{figure}[!t]
	\centering
		\includegraphics[width=0.98\textwidth]{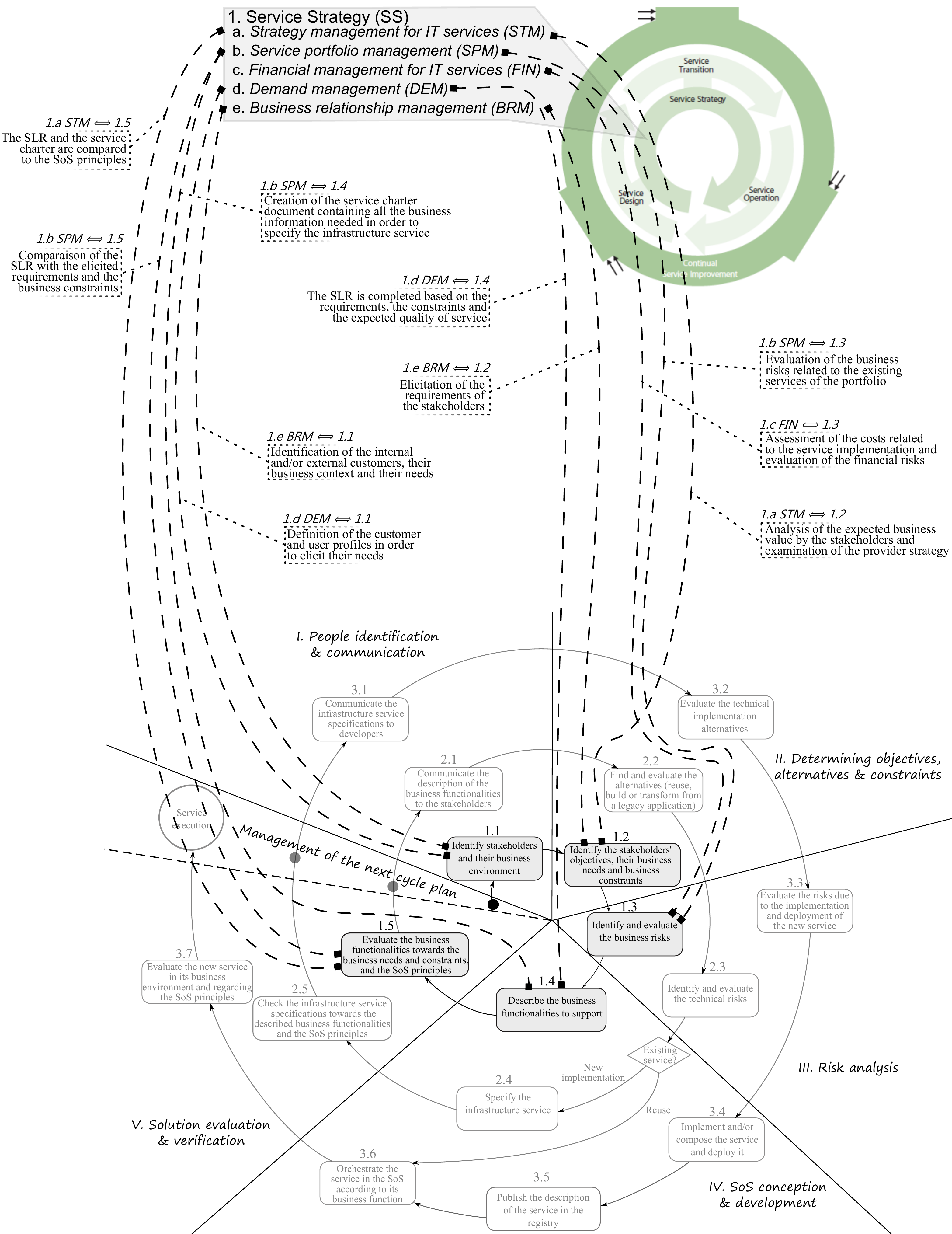}
	\caption{Illustration of the alignment between the first cycle of the \xc{s}o\xc{s} provisioning model and the \xc{itil v.}$3$ processes}
	\label{fig:Illustration1cycleSOSprovAndITIL}
\end{figure}

\begin{figure}[!t]
	\centering
		\includegraphics[width=1.02\textwidth]{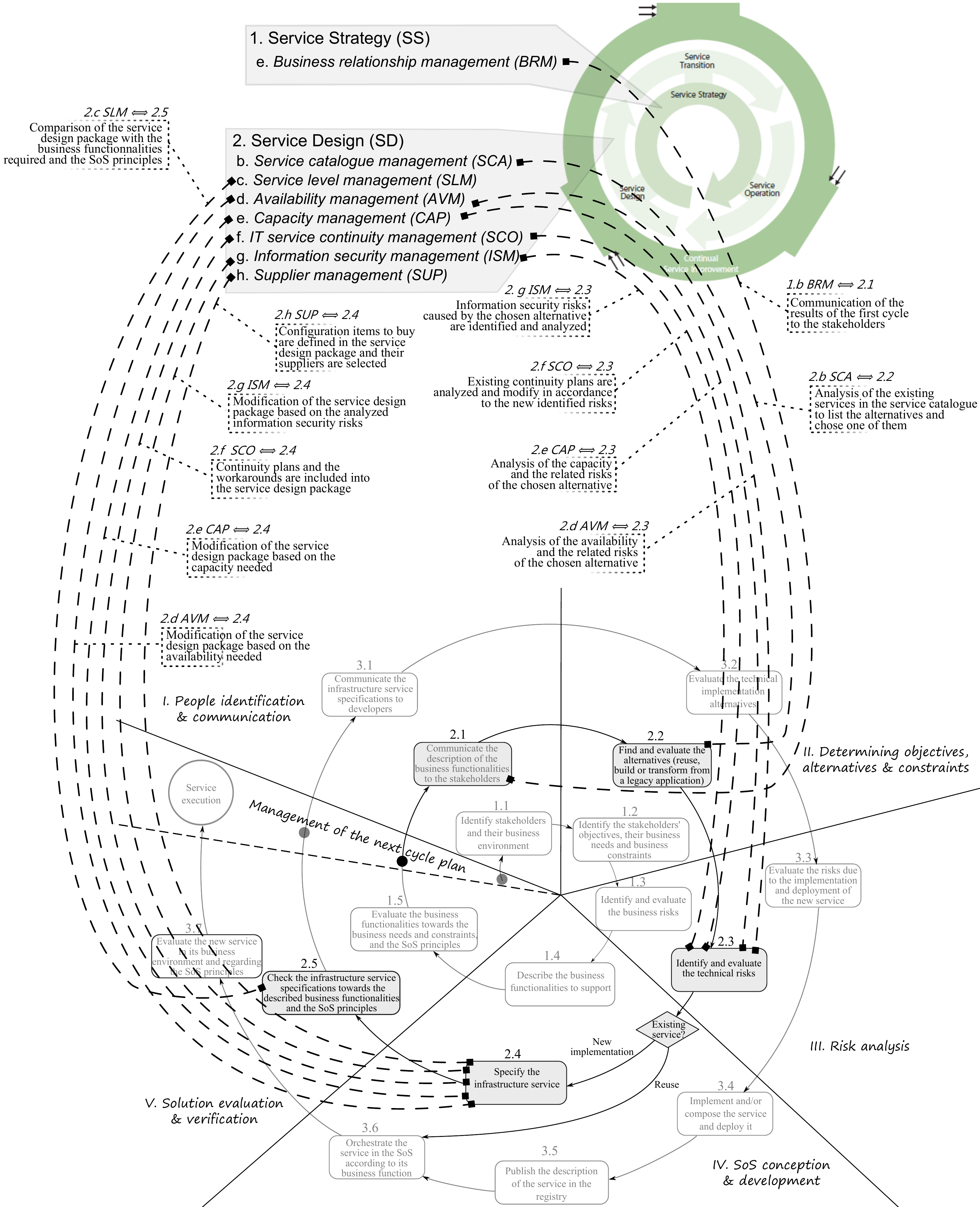}
	\caption{Illustration of the alignment between the second cycle of the \xc{s}o\xc{s} provisioning model and the \xc{itil v.}$3$ processes}
	\label{fig:Illustration2cycleSOSprovAndITIL}
\end{figure}

\begin{figure}[!t]
	\centering
		\includegraphics[width=1.00\textwidth]{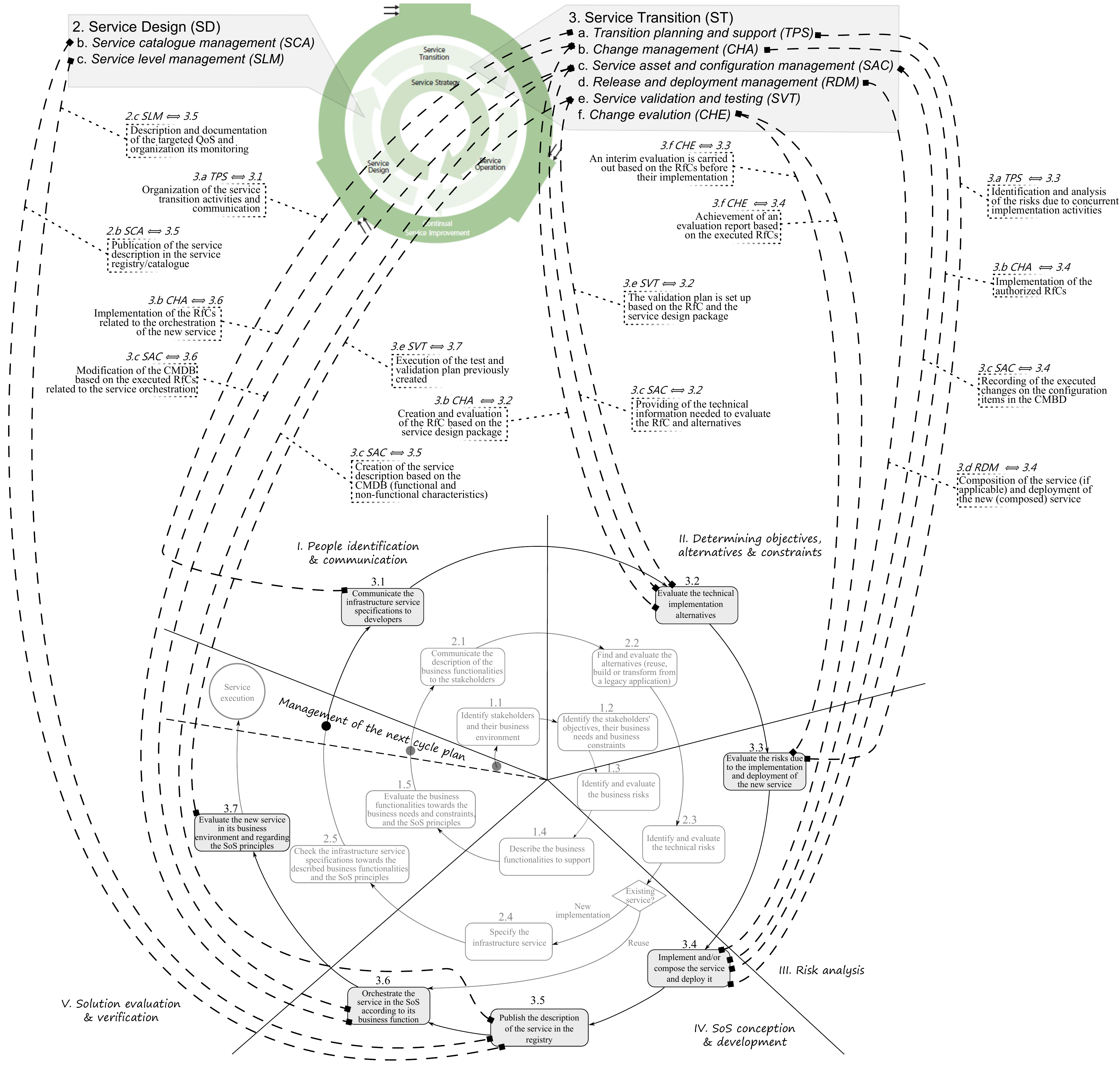}
	\caption{Illustration of the alignment between the third cycle of the \xc{s}o\xc{s} provisioning model and the \xc{itil v.}$3$ processes}
	\label{fig:Illustration3cycleSOSprovAndITIL}
\end{figure}

Figures~\ref{fig:Illustration1cycleSOSprovAndITIL},~\ref{fig:Illustration2cycleSOSprovAndITIL} and~\ref{fig:Illustration3cycleSOSprovAndITIL} illustrate the alignment between the \xc{s}o\xc{s} provisioning model and \xc{itil v.}3. Their main objective is to ease the understanding and the use of the alignment proposed. Each illustration corresponds to one cycle of the \xc{s}o\xc{s} provisioning model.

After the provisioning of a new service in the \xc{s}o\xc{s}, the next stage is the service execution (see the end of the third cycle in Figure~\ref{fig:SOSprovisioningLifeCycle}). This corresponds to the use of the service functionalities when the business processes are carried out. This will be supported thanks to the service operation phase of \xc{itil} (see~\S\ref{presentationITIL}). Note this phase and its alignment with the service-oriented paradigm is not in the scope of this work, but this issue deserves further investigations. The possible relations between the improvement of the service provided, described in \xc{itil} by the Continual Service Improvement phase, is also out of the scope of this work although this analysis should be very interesting.

A last remark concerns the Knowledge management process (3.g \xc{kno}) that supports all the \xc{s}o\xc{s} provisioning steps detailed previously. Indeed, this process aims at sharing and providing all the types of knowledge needed --i.e., the perspective, the experience, the information and the ideas-- to organize the tasks and to reach their objectives. The ``inputs to knowledge management include all knowledge, information and data used by the service provider as well as relevant business data''~\cite[\S 4.7.6.2]{itilST}.

\section{Validation Framework\label{validationFrameworkProposed}}
Along with the alignment between the steps of the \xc{s}o\xc{s} provisioning model and the \xc{itil v.}3 processes, we shape a validation framework which should be applied in order to validate this work. It must be used in a real and natural context in the form of a field study --also named case study or observational study--. It appears to be the best solution in our situation~\cite{DBLP:journals/ese/RunesonH09}. This validation framework shapes the analysis of feedbacks collected from software engineering teams in charge of a service-oriented architecture and working for a company respecting the \xc{itil v.}3 best practices.

This section is organized as follows. First, we broadly describe the validation framework proposed~(\S\ref{ValidFrameworkCharacteristics}). Then, the preparation of the field study~(\S\ref{dataCollectionPrepa}) and the data collection~(\S\ref{DataCollection}) are addressed. After some remarks concerning the analysis of the information collected~(\S\ref{EvidenceAnalysis}), we detail the expected structure of the report~(\S\ref{ResultsReporting}). Note these subsections correspond to the process steps needed to report a case study in software engineering contexts~\cite{DBLP:journals/ese/RunesonH09}. In the last part~(\S\ref{ThreatsValidity}), we mention the expected threats to validity.

\subsection{Characteristics of the Validation Framework Proposed\label{ValidFrameworkCharacteristics}}
The main objective of this validation framework is to evaluate the effects of the work proposed on the software engineering teams which use it and, to a lesser extent, on the whole organization. The research question is: ``How the use of the alignment between the \xc{itil v.}$3$ processes and the \xc{s}o\xc{s} provisioning steps can ease the daily work of software engineering teams?''. Although the conduction of field studies is roughly flexible regarding the data collection and analysis, their objective and their underlying research question(s) cannot be modified from the beginning to their end. At the end of the field study, we expect to draw some hypotheses concerning the use of the alignment between \xc{itil v.}3 and the \xc{s}o\xc{s} provisioning model. In subsequent field research, these hypotheses should be confirmed in order to increase the external validity of the first results obtained.

They are several types of field study: exploratory (finding out and generating hypotheses and ideas for new research), descriptive (describing a phenomenon), explanatory (researching an explanation of facts or problems), confirmatory (validating and confirming a theory, a relation or a hypothesis) and emancipatory (improving some aspects of already studied situations, relations, hypotheses or phenomenons)~\cite{robson2011realWorldResearch,easterbrook2008selecting}.
\\The field study which is being described can be categorized as an exploratory study --also called case study. Indeed, its objective is to know how the alignment between \xc{itil v.}3 and the \xc{s}o\xc{s} provisioning helps the software engineering teams. It should enable the generation of hypotheses and ideas from the use of the proposed material. As detailed below, the structure of the field study is based on a positivist and qualitative point of view.

The following characteristics are needed in the context of the field study. Of course, the field selection mainly depends on its availability~\cite{DBLP:journals/tse/SjobergHHKKLR05}. Therefore, some of the undermentioned characteristics might vary slightly. First, the software engineering team works in a company that has implemented the \xc{itil v.}$3$ best practices and that owns at least one \xc{s}o\xc{s}. Some members of the software engineering team should have a medium to high experience in an \xc{itil} company and/or in the implementation of \xc{it} services. They have to accept to take part in the field study. Note this is very important that the researchers have some clear and signed agreements with the company leaders and with each member of the software engineering team interviewed. A second need concerns the projects. The company should have a relatively broad list of future projects related to the implementation or the modification of \xc{it} services.

In order to collect the data, we recommend to use a direct method which mainly consists of interviewing the participants. We favour the use of semi-structured interviews\footnote{Other direct methods are structured interviews~\cite{lincoln1985naturalistic}, unstructured interviews~\cite{becker2012understanding} or focus group~\cite{krueger2009focus}. Structured interviews contain a list of predetermined questions with no possible follow-up questions. They are mostly used for clarification purpose. On the opposite side, unstructured interviews are performed without any organization and do not reflect any predetermined ideas. The third possible technique, i.e., organizing focus group, should be avoided in our context. Indeed, in case of specific social or cultural norms --such as hierarchic relations-- the focus group participant may stay silent about significant aspects in the scope of the research or they may repeat the main opinion of the group~\cite{kitzinger1995qualitative}. We would avoid that managers or project leaders influence the declarations of their subordinates. Therefore, the semi-structured interviews consisting of several key questions covering the area to explore seem to be the right way to conduct the interviews. A second advantage is to let the interviewer explore in more details an idea or a given answer with follow-up questions such as ``Can you tell me more about that?''.}.
Its most significant advantage is to be sure that all important and relevant topics for the research are tackled, also in the scope of small-scale research~\cite{drever1995using}. We create a semi-structured interview guide\footnote{All the questions should be addressed but not necessarily in the same order.}. The interview guide covers three main themes: the profile of the interviewee, the practical use of the \xc{itil v.}3/\xc{s}o\xc{s} provisioning alignment, and its potential utility.

\begin{enumerate}
	\item The profile of the interviewee: \zi{the objectives of this theme are to know the experience of the interviewees, to identify the links between them and to put their jobs and tasks in relation with the two others themes tackled}.
		\begin{itemize}
			\item Could you briefly describe your experience and your skills mobilized in the scope of your work?
			\item Could you detail your current role and your tasks in the company?
			\item Could you detail your role and your involvement during the last projects leading to the \xc{s}o\xc{s} provisioning with new services?
		\end{itemize}
		
	\item The practical use of the \xc{itil v.}3/\xc{s}o\xc{s} alignment: \zi{the objective is to understand how they use the illustrations of the alignment provided in their daily work as well as in the organization and the management of the projects}.
		\begin{itemize}
			\item What did you remember concerning the presentation of the \xc{s}o\xc{s} provisioning model and its alignment with \xc{itil v.}3?
			\item Could you describe how you used the material provided before and during the projects?
			\item In comparison to the project organization, what kind of information did you receive in addition to the material provided?
			\item Based on your experience, what could have been done differently concerning the use of the material provided?
			\item How do you think the alignment between \xc{itil v.}3 and the \xc{s}o\xc{s} provisioning model and its use could be improved?
		\end{itemize}
	
	\item The potential utility of the \xc{itil v.}3/\xc{s}o\xc{s} alignment: \zi{the objective is to understand if the \xc{itil v.}3/\xc{s}o\xc{s} alignment helps the actors of the project and who are these actors}.
		\begin{itemize}
			\item What were your feelings when the project was announced by the managers of your company?
			\item What were your feelings when the project started?
			\item What were your feelings when the alignment material was presented and explained?
			\item Could you explain your opinion about the illustrations of the \xc{s}o\xc{s} provisioning model and its alignment with \xc{itil v.}3?
			\item Could you give and, if possible, explain the strengths and the weaknesses of the material provided?
			\item What is the main sentiment among your colleagues about the use of this material?
		\end{itemize}
\end{enumerate}
Before the first question, the interviewees should be encouraged to response in details. The confidentiality of the content of the discussions should be guaranteed. Only aggregated and unnamed results and analysis may be disclosed. Note these questions may be sent to the interviewees approximately one week before the meeting in order to allow them to be prepared and, possibly, less anxious about the exact content of the discussion.
\\This interview should end with this last and broad question: ``Do you want to add or change something to the discussion?''. As explained in~\cite{DBLP:conf/metrics/HoveA05}, we recommend to provide the visual illustrations of the alignment to the interviewees before the beginning of the interview. These illustrations could be the illustration of the \xc{s}o\xc{s} provisioning model and its alignment with \xc{itil v.}3 such as the one shown in Figure~\ref{fig:Illustration1cycleSOSprovAndITIL}. The interviewees can use these documents to illustrate their words.

\subsection{Preparation of the Field Study\label{dataCollectionPrepa}}
At the beginning of the field study, the \xc{s}o\xc{s} provisioning model and its alignment with \xc{itil v.}3 should be presented to the software engineering team and company leaders. The available material is, for instance, Figure~\ref{fig:Illustration1cycleSOSprovAndITIL} which illustrates the alignment of the first cycle of the \xc{s}o\xc{s} provisioning model with the \xc{itil v.}3 processes. The main objectives of the research as well as the research procedure followed should also be explained to the participants. After this presentation, we recommend to let them work alone in their projects.
\\When one or several projects leading to the implementation of \xc{it} services end, the data collection can be organized (see the next section).

\subsection{Achievement of the Field Study and Data Collection\label{DataCollection}}
Based on the interview guide described in~\S\ref{ValidFrameworkCharacteristics}, one or, preferably, two researcher(s) can conduct the interviews. It is important to interview different roles who take part in the projects. As an example, the roles of project manager, change manager, business analyst, programmer, applications owner and tester are relevant. This selection can be made based on the project documentation, the CVs, the hierarchy of the company and so on. The triangulation of the data sources helps to draw strong and receivable conclusions. The interviews should end when no more relevant information are gotten from the last interviews.
\\We recommend to refer to the work of Hove \& Anda~\cite{DBLP:conf/metrics/HoveA05} who gives very sound advice to conduct semi-structured interviews in the scope of qualitative research in software engineering. In particular, they advise to record the interviews in order to ease their transcriptions and analysis. They also advise to take note about the non-verbal communication of the interviewees, e.g., the silences, grimaces, laughing and so on. The analysis of the non-verbal communication is often relevant. Indeed, there are different ways to remember facts and experiences because of, i.a., the different backgrounds, know-how, knowledge or responsibilities. In most of the cases, this is partially disclosed by the face and corporal expressions.

\subsection{Analysis of the Evidence Collected\label{EvidenceAnalysis}}
Once the first interviews conducted and transcribed, we foster a qualitative analysis since the data collection method proposed is the  semi-structured interviews. This analysis should be conducted in parallel with the interviews. Indeed, first discussions and data collections often raise new issues and insights in exploratory field studies. Accordingly, it is essential to adapt the interview guide by adding or modifying questions.
\\A second significant point concerns the number of researchers who carry out the data analysis. Having multiple and independent researchers conducting the analysis decrease the risks of having bias. At the end of the first phase of the analysis process, the preliminary analysis achieved by each researcher are compared and merged in order to report a common view of the field study.

Possible techniques to conduct the analysis are, e.g., the grounded theory~\cite{discoveryGroundedTheory} or the cross-case analysis~\cite{eisenhardt1989building} if multiple field experiments can be conducted for the same study --a short and practical description of these two techniques can be found in~\cite{DBLP:journals/tse/Seaman99}. These analysis techniques should help to generate the hypotheses, derived from the results, that will be presented in the final report (see~\S\ref{ResultsReporting}). Researchers can use software solutions such as \xc{nv}ivo (commercial software) or \xc{catma} (open source).

\subsection{Reporting of the Results\label{ResultsReporting}}
The main research results expected are  hypotheses about, i.a., the modification in the project management, in the performance of the software engineering team, in the quality of the \xc{it} service delivered or in the usability of the \xc{itil v.}3/\xc{s}o\xc{s} alignment. The properties of the studied software engineering team and its company together with the characteristics of the interviewed people are significant aspects to report. As an example, their function, their experience and knowledge in \xc{itil} and \xc{s}o\xc{s}, and their involvement in the next software engineering projects are relevant pieces of information.
\\The sequences of actions initiated by the researchers should be clearly described (mainly the \zi{what}, \zi{whom} and \zi{how} questions). This is important for a subsequent validation of the hypotheses and for the comparison with similar field studies. 

Some further guidelines to report field experiments can be found in~\cite{DBLP:conf/isese/JedlitschkaP05,yin2009case}. Note that the software engineering team which has been interviewed must be kept informed of the research results.

\subsection{Threats to Validity\label{ThreatsValidity}}
There are four main factors to consider when dealing with validity in qualitative research: credibility, transferability, dependability and confirmability~\cite{lincoln1985naturalistic}. Concerning the credibility factor, i.e., are the results reported believable from the viewpoint of the research stakeholders, a clear approval of the research results presented to the software engineering team and to the company leaders is essential. The transferability factor, i.e., the degree to which the results reported can be generalized to other contexts, will be very difficult to fully satisfy. Indeed, the research design focuses on a single software engineering team. However, the main objective is first to generate hypotheses about the use of the alignment between the \xc{s}o\xc{s} provisioning model and \xc{itil v.}3. A more significant focus should be given to the transferability factor when these hypotheses will be tested and validated in other software engineering teams. Regarding the dependability, it corresponds to the degree to which the data collection and analysis depend on the researchers conducting the research work. The way the data were collected, and how they are coded and analysed should be very clear in order to allow other people to check and redo the research procedure followed. The last validity factor is the confirmability. It covers the degree to which the results reported could be corroborated and validated by other researchers. The confirmability could be reached if several researchers reproduce the same study in other contexts with similar results.
\\Note some authors propose strategies to verify the validity of qualitative research. As a relevant example, see the work of Morse et al.~\cite{morse2002verification}.  

\section{Related Work\label{relatedWork}}
Several works address and associate the service-oriented paradigm and \xc{itil}. Some of them establish different kinds of relations between them~(\S\ref{SOAitilLinks}). Other works concern the \xc{s}o\xc{s} governance which is shaped thanks to some \xc{itil} best practices~(\S\ref{soaGovernanceItil}).

\subsection{Establishment of Relations Between the ITIL Phases and Processes and the Service-Oriented Paradigm\label{SOAitilLinks}}
The relations proposed between \xc{itil} and the service-oriented paradigm are often based on the organizational concepts of \xc{itil}, and on the \xc{s}o\xc{s} concepts and implementation steps. First initiatives combining the \xc{s}o\xc{s} with management practices and organizational aspects focus on the service-oriented software engineering (see~\cite{karhunen2005service} for more details). In the scope of this work, we only consider the works referencing the \xc{itil} framework.

In~\cite{DBLP:conf/sac/BraunW07}, the authors propose a meta model of an enterprise service based on the service concept of \xc{itil v.}2 and of the service-oriented paradigm. They do not tackle the possible relations between the \xc{itil} processes and the activities of the \xc{s}o\xc{s} implementation and provisioning.
\\Other works such as~\cite{JTAITceaFramework} use \xc{itil v.}3 concepts to build a service-oriented and organizational framework. But they do not align the processes of \xc{itil} with processes or activities of an \xc{s}o\xc{s} implementation or provisioning methodology.

In~\cite{DBLP:conf/cmg/WaringSD05}, the author favors the use of an \xc{soa} integrated with \xc{itil} in order to improve the agility of \xc{it} in organizations. This integration helps to relate the management of \xc{it} service with their supportive technical layers, which are assimilated to the \xc{soa} components. They use the second version of \xc{itil} --the third one was not yet finished when the work has been published. Most of the \xc{itil v.}2 processes are related to the \xc{soa} concepts. A particular attention is paid to the \xc{cmdb} management and the operational activities, i.e., the management of the services configuration, the incidents, the requests and the problems. A second work shares a similar objective, i.e., improving the \xc{it} agility by combining an \xc{soa} and \xc{itil}~\cite{DBLP:journals/ijitm/IzzaI10}. Its authors claim for a clear distinction between the \xc{soa} concepts and the \xc{itil} concepts\footnote{In~\cite{DBLP:journals/ijitm/IzzaI10}, the authors prefer the notion of \xc{itsm} for managing the \xc{it} services. Given that the most popular \xc{itsm} framework is \xc{itil}, we only refer to \xc{itil} in our related work.} although they underline some connection between these two organizational domains.
\\Compared to~\cite{DBLP:conf/cmg/WaringSD05} and~\cite{DBLP:journals/ijitm/IzzaI10}, our work goes one step further by aligning the steps of a service-oriented development methodology and the \xc{itil} process. Of course, the scope of our work is narrowed since we only focus on the creation of services. The operational management of the built services is left for future work. A third work is very close to this idea. In~\cite{ganek2007overview}, the authors describe the technical platform used by \xc{ibm} to manage the services in a \xc{s}o\xc{s}. They clearly refer to \xc{itil} best practices and principles. Of course, the use of the alignment between the \xc{s}o\xc{s} components and \xc{itil} is only possible if the \xc{ibm} software tools are purchased. Moreover, these relations are not described neither justified.

Other works propose to associate the service-oriented paradigm with an organizational framework which is not \xc{itil}. For instance, Li et al. design a high level organizational framework and structure which is then compared and aligned to the \xc{soa} implementation~\cite{DBLP:conf/enase/LiZO11}. They consider that the \xc{soa} is a mirror of real organizations. This choice is motivated by the need for a technology independent framework. We meet this requirement by using \xc{itil} as the reference organizational framework which is independent of any specific technologies. The detailed description of \xc{itil} is an advantage compared to the use of a high level and less described organizational framework. A second example is the Service-Oriented Analysis and Design method (\xc{soad})~\cite{zimmermann2004elements}. The authors cover the business and organizational layers in their model, but without reference to a detailed organizational framework.

\subsection{Service-Oriented Governance Based on ITIL Concepts\label{soaGovernanceItil}}
Broadly speaking, the governance of the service-oriented architectures aims at coordinating the \xc{it} decisions and the business goals of organizations. \xc{itil} is one of the reference governance models used.
\\In~\cite{DBLP:conf/iceei/SusantiS11}, the authors propose some links between an service-oriented governance model and the \xc{it} governance model of the \xc{itil v.}3 framework. These links are concentrated on the service concept and its life cycle, although the reference definition of the service notion is not given. In order to detail them, they mix up the life cycle of a service, of the service provisioning and of an \xc{soa}. A very similar work~\cite{mapping6phaseSOAandITIL} which extends the paper~\cite{DBLP:conf/iceei/SusantiS11} does not clarify the issues raised.
\\In~\cite{DBLP:conf/cason/FilhoA12,filhogovernance}, the authors propose a consolidate approach of organization management that addresses the main issues as the organization governance. They called it the CommonGov approach. Based on the main recent governance models proposed in the literature, they create a service-oriented governance based on some \xc{soa} processes. One of the models used is \xc{itil v.}3.
\\In~\cite{DBLP:conf/apscc/Xian-PengBR12}, the authors propose \xc{itil v.3} as the \xc{s}o\xc{s} governance solution. The governance of services with \xc{itil} is globally well described. Our work concentrates on the service creation while~\cite{DBLP:conf/apscc/Xian-PengBR12} focuses on its overall management.
\\Other papers, e.g.,~\cite{DBLP:conf/sac/SchepersIE08,DBLP:conf/edoc/KohlbornKR09}, propose service-oriented governance models without any reference to \xc{itil}. Seeing that \xc{itil} is currently one of the most known and used organizational framework for \xc{it} services providers~\cite{marrone2014service} --\xc{itil v.}3 includes all the aspects related to the organizational governance--, we argue that creating a new governance model is unnecessary compared to the use of an established model applied in many companies around the world.

\section{Conclusion and Future Work\label{CCL}}
The business processes of organizations providing \xc{it} services are regularly based on the \xc{itil} framework. The \zi{service orientation}, which is applied at the organizational level thanks to the \xc{itil} best practices, should also be followed at the \xc{is} level. The service-oriented paradigm is a solution to achieve this objective. Although this paradigm and \xc{itil} both promote a service orientation, it is not clear how to align these two levels of an organization. In other words, the organizational processes of \xc{itil v.}3 do not clearly correspond to the activities of the existing models followed to implement and provision new services.
\\In this work, we first detail a service oriented development methodology into the Spiral Model, and then we align it with the \xc{itil v.}3 processes. Among others, we define the inputs and outputs in each identified relations between the \xc{s}o\xc{s} provisioning model and \xc{itil v.}3. This work makes clear and eases the understanding of the alignment between the service implementation life cycle and the \xc{itil v.}3 life cycle. The illustrations of the proposed alignment should help the collaborators and managers to organize their work and relationships according to the service creation methodology, and in accordance with the \xc{itil} best practices and the principles of the service-oriented paradigm.
\\Along with these contributions, we shape a validation framework which can be used to generate hypotheses about the use of this work in a real environment. Once this exploratory study achieved, a second phase of the validation work should be the hypotheses assessment. Of course, one of our future work is to achieve an exploratory study based on the validation framework proposed.

Another main future work lies in the analysis of the service execution in order to identify the relations with the \xc{itil v.}3 phases and processes.

\bibliographystyle{splncs}
\bibliography{BibTex-IESS-TP}

\end{document}